\documentclass[ 11pt,review]{elsarticle}

\usepackage[utf8]{inputenc}
\usepackage{graphicx}

\usepackage{hyperref}
\usepackage{booktabs}

\usepackage{subfig}

\usepackage{amsmath}

\usepackage[hyperpageref]{backref} 
 \biboptions{sort,compress,super}

\usepackage{multirow}
\usepackage{graphicx}
\usepackage{xcolor}
\usepackage[export]{adjustbox}
\usepackage{float}

\usepackage{subfig}
\usepackage{tikz}

\usepackage{etoolbox}
\patchcmd{\pprintMaketitle}
  {\fi\hrule}% the second rule
  {\fi\ifvoid\extrainfobox\else\unvbox\extrainfobox\par\vskip10pt\fi\hrule}
  {}{}

\newenvironment{extrainfo}
  {\global\setbox\extrainfobox=\vbox\bgroup\parindent=0pt }
  {\egroup}
\newsavebox\extrainfobox

\newcommand{\beginsupplement}{%
        \setcounter{table}{0}
        \renewcommand{\thetable}{S\arabic{table}}%
        \setcounter{figure}{0}
        \renewcommand{\thefigure}{S\arabic{figure}}%
     }

\usepackage{verbatim}

\begin{document}
\begin{frontmatter}
% %Don't count these!
% %TC:ignore
% \quickwordcount{main}
% \quickcharcount{main}
% \detailtexcount{main}
% %TC:endignore

\title{Deep Learning for Predicting Progression of Patellofemoral Osteoarthritis Based on Lateral Knee Radiographs,  Demographic Data and Symptomatic Assessments}

\author[1]{Neslihan Bayramoglu\corref{cor1}%
\fnref{fn1}}
\author[2]{Martin Englund}
\author[3]{Ida K. Haugen}
\author[4]{Muneaki Ishijima}
\author[1,5]{Simo Saarakkala}

\begin{extrainfo}
\textbf{Running title:} Predicting Patellofemoral Osteoarthritis Progression
\newpage
\end{extrainfo}

\cortext[cor1]{Corresponding author}
\fntext[fn1]{POB 5000, FI-90014, Oulu, Finland}

\address[1]{Research Unit of Health Sciences and Technology, University of Oulu, Finland}
\address[2]{Orthopaedics, Department of Clinical Sciences Lund Faculty of Medicine, Lund University}
\address[3]{Center for Treatment of Rheumatic and Musculoskeletal Diseases (REMEDY), Diakonhjemmet Hospital, Oslo, Norway}
\address[4]{Department of Orthopaedics, Faculty of Medicine, Juntendo University} 
\address[5]{Department of Diagnostic Radiology, Oulu University Hospital, Oulu, Finland}

\begin{abstract}
\small

\noindent \textbf{Objective:}
%Patellofemoral osteoarthritis (PFOA) is a common and debilitating condition that can greatly affect quality of life. However, PFOA has been investigated much less than tibiofemoral OA. Accurately predicting the radiographic progression of PFOA is crucial for early intervention and effective management of the disease. 
In this study, we propose a novel framework that utilizes deep learning and attention mechanisms to predict the radiographic progression of patellofemoral osteoarthritis (PFOA) over a period of seven years. 
%The framework takes into account lateral radiographs, demographic data, and symptom assessments (WOMAC), allowing for a more comprehensive and accurate prediction of disease progression.

\noindent \textbf{Design:} 
This study included subjects (1832 subjects, 3276 knees) from the baseline of the Multicenter Osteoarthritis Study (MOST).
%with the exclusion of those whose PFOA status could not be assessed over the following seven years due to missing data. 
Patellofemoral joint regions-of-interest were identified using an automated landmark detection tool (BoneFinder) on lateral knee X-rays.
An end-to-end deep learning method was developed for predicting PFOA progression based on imaging data in a 5-fold cross-validation setting. To evaluate the performance of the models, a set of baselines based on known risk factors were developed and analyzed using gradient boosting machine (GBM).
Risk factors included age, sex, %body mass index (BMI)
BMI and %McMaster Universities Arthritis Index (WOMAC) 
WOMAC score, and the radiographic osteoarthritis stage of the tibiofemoral joint (KL score).
Finally, to increase predictive power, we trained an ensemble model using both imaging and clinical data.
%We employed the area under the ROC curve (AUC-ROC),  the area under the PR curve (average precision, AP), and Brier score for the assessment of the models' performance.

\noindent \textbf{Results:}
Among the individual models, the performance of our deep convolutional neural network attention model %(VGG-16-Attn, Model5) 
achieved the best performance with an AUC of 0.856 and AP of 0.431; slightly outperforming the deep learning approach without attention (AUC=0.832, AP= 0.4) and the best performing reference GBM model (AUC=0.767, AP= 0.334).
The inclusion of imaging data and clinical variables in an ensemble model allowed statistically more powerful prediction of PFOA progression (AUC = 0.865, AP=0.447), although the clinical significance of this minor performance gain remains unknown. 
% The stacked ensemble model can accurately distinguish between patients who are likely to experience PFOA in the future and those who are not. 
%We also observed that high BMI, female sex, and OA in the tibiofemoral joint increase the risk of PFOA development, and subjects with higher WOMAC scores are more likely to develop PFOA in the future. 
The spatial attention module improved the predictive performance of the backbone model, and the visual interpretation of attention maps focused on the joint space and the regions where osteophytes typically occur.

\noindent \textbf{Conclusion:}
This study demonstrated the potential of machine learning models to predict the progression of PFOA using imaging and clinical variables. These models could be used to identify patients who are at high risk of progression and prioritize them for new treatments. However, even though the accuracy of the models were excellent in this study using the MOST dataset, they should be still validated using external patient cohorts in the future.
\end{abstract}


\end{frontmatter}
\section{Introduction}

Knee osteoarthritis (OA) is the most prevalent chronic joint disorder that involves degeneration and loss of articular cartilage along with bony changes.
High age and body mass index (BMI) are strong risk factors for knee OA \cite{lankhorst2017incidence}. 
Structural knee OA often leads to significant pain, stiffness, disability, and reduced quality of life for affected individuals\cite{duncan2008pain}. 
Current understanding of OA disease process is inadequate and, consequently, there is a lack of disease-modifying medical treatments.
As a result, knee OA continues to impose a significant burden on individuals and society \cite{crossley2011patellofemoral}.

Although the patellofemoral (PF) joint is an important source of symptoms in knee OA,
the majority of the research on knee OA has focused on tibiofemoral (TF) joint of the knee \cite{duncan2011incidence, duncan2009does}.
Patellofemoral OA (PFOA) can be caused by a number of factors, including previous injury to the knee, inflammation, biomechanical abnormalities, overuse of joint, obesity, and genetic predisposition \cite{kim2012patellofemoral, crossley2011patellofemoral}. 
Symptoms often include anterior knee pain, especially when kneeling and squatting, as well as swelling and a grinding or popping sensation when moving the knee (crepitus)\cite{van2018international}.
As the importance of the PF joint in OA is increasingly acknowledged, the number of studies into it has been increasing \cite{kobayashi2016prevalence, macri2017patellofemoral, crossley2011patellofemoral}. 
Still, more research is needed \cite{crossley2011patellofemoral}.

Non-invasive imaging techniques play a crucial role in diagnosing and monitoring PFOA.
Without imaging, a confident diagnosis will seldom be possible for PFOA \cite{peat2012clinical}.
X-ray imaging is one of the primary diagnostic tools because of its low cost and wide availability. 
Although radiography does not allow to visualize soft tissues, changes in the joint space and bone structure can be well depicted from X-rays. 
Several imaging biomarkers such as the narrowing of the joint space, bony spurs, malalignment of the patella, bone sclerosis, and cysts are associated with PFOA\cite{de2016radiographic, macri2017patellofemoral}

In recent years, machine learning (ML) techniques have emerged as promising tools to aid in the diagnosis of PFOA from X-ray images \cite{bayramoglu2021pfoa, bayramoglu2022machine}.
Both early diagnosis and prediction of disease progression might be critical in the management and intervention of PFOA.
However, accurate and timely identification of PFOA progression based on X-ray images can be challenging due to the complexity of the disease and the variability of knee imaging. To date, there are no published studies using ML-based models for prediction of PFOA development or progression in the future from imaging data.

In this study, we introduced a deep learning based framework to predict radiographic progression of PFOA over a 7-year period from lateral radiographs, demographic data and symptom assessments (clinical data).
We leveraged attention mechanism in our deep learning framework and proposed an end-to-end solution via a trainable
attention module.
The results of this study have the potential to improve the early diagnosis and treatment of PFOA, ultimately leading to improved patient outcomes and quality of life.

\section{Materials and Methods}
\begin{figure}[H]
 \captionsetup[subfigure]{justification=centering}
    \centerline{
    \begin{adjustbox}{color = blue, varwidth=1.3\textwidth,margin=3pt 1pt 3pt 0,frame=0.5pt }
   {\hspace{-0.0cm}\includegraphics[trim={0cm 0.0cm 0 0.0cm},clip, width=0.99\textwidth]{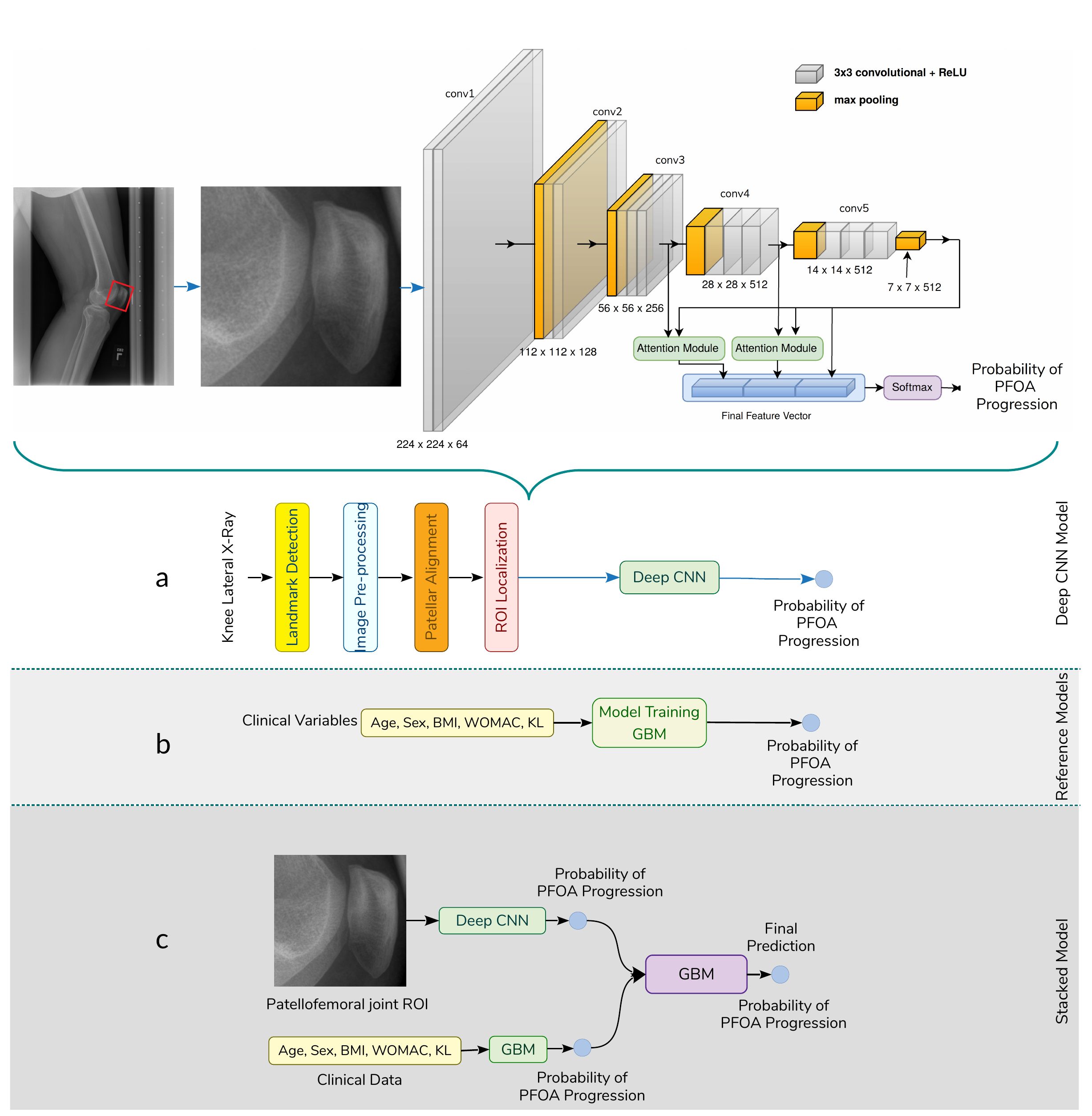}}

    \caption{ a) Illustration of the workflow of our approach. The localization and alignment of patellofemoral (PF) joint in lateral knee X-rays were performed based on the anatomical landmarks of patellar bone (BoneFinder).   Intensity normalization was then applied.
    Finally,  each lateral knee  was  rotated  in  order  to  have  an  aligned  patella.  
After localizing PF joint ROI, a deep convolutional neural network (CNN) model was used for predicting the progression of patellofemoral osteoarthritis (PFOA).
    b) For comparison, a separate machine learning model (gradient boosting machine (GBM)) was trained  based on clinical variables including age, sex, body mass index (BMI), the total Western Ontario and McMaster Universities Arthritis Index (WOMAC) score, and Kellgren and Lawrence (KL) score of the tibiofemoral joint.
    We used a stratified subject-wise 5-fold cross validation setting to measure the performance of all the models. 
    c) In addition to these individual models, we fused the predictions from these models in a second layer GBM model to  improve the overall prediction performance.
    }
    \label{fig:flowchart}
    \end{adjustbox}
    }
\end{figure}

Figure \ref{fig:flowchart} shows the overall pipeline of our study.
We first located patellar landmarks using BoneFinder software\cite{lindner2013fully}
(Figure \ref{fig:roi}).
Those anatomical landmarks were then used to align patellar bone constantly across the knees eliminating rotation variance.

The image preprocessing step involved normalizing intensity using global contrast normalization and truncating the histogram between the $5^{th}$ and $99^{th}$ percentiles. Subsequently, we used patellar landmarks to locate the patellofemoral joint regions of interest (PFJROI) in lateral knee radiographs. To ensure a similar view with left knee images, the right knee ROI images were horizontally flipped. We then utilized a deep convolutional neural network (CNN) to predict PFOA progression within 7 years. Additionally, we trained a machine learning model (GBM \cite{friedman2001greedy}) on clinical features as a reference method for comparison with the proposed approach. Finally, to increase predictive power, we trained an ensemble model using both imaging and clinical data.

%TC:ignore
%TC:endignore

\subsection{Data}
We used the data from the Multicenter Osteoarthritis Study public use datasets (MOST, \href{http://most.ucsf.edu}{http://most.ucsf.edu}).
MOST is a longitudinal observational study that aims to identify factors affecting the occurrence and progression of OA.
The study enrolled 3,026 participants aged 50–79 years who either had radiographic knee OA or were at high risk for developing the disease.
The participants has been followed 84 months where clinical assessments were conducted and radiological data were collected.
In the study, semiflexed lateral view radiographs were acquired according to a standardized protocol.
Knee radiographs were evaluated from the baseline to 15, 30, 60 and 84-month follow-up visits.
In this study, we employed lateral radiographs acquired at the baseline visit from both left and right legs that includes 3276 knees (1832 subjects) which did not have PFOA at the time of first examination.
The number of progressed knees that developed PFOA is 403 ($12\%$) and the number of knees which did not develop PFOA is 2873 ($88\%$).
Selected knees must have had PFOA assessments from lateral radiographs and KL grades from posteroanterior (PA) radiographs, all performed at the baseline.  
Among those ones, we selected knees only whose patellofemoral OA status within the following 7 years can be assessed (progressor vs non-progressor).
For example, participants who dropped out from the study before the last follow-up timepoint and had not developed PFOA at the previous time points were excluded.
See supplementary material for subject flow diagram and demographics.

In the MOST public use datasets, radiographic PFOA is defined from lateral view radiographs as follows: Osteophyte score $\geq$ 2 or the joint space narrowing (JSN) score is $\geq1$ plus any osteophyte, sclerosis or cysts $\geq1$ in the PF joint (grades 0–3; 0=normal, 1=mild, 2=moderate, 3=severe).
Unlike tibiofemoral joint OA assessment (KL grading ranging from 0 to 4), in the PF joint, OA was described either present or absent lacking a severity grading. 
In this study, the term ``progression" refers to both progression of existing OA and development of OA in previously non-affected PF joints (incidence)
%Based on this, we used both ''development" and ''progression" terminology in this study.
For example, knees which showed minor signs of PFOA (e.g. osteophyte score=1) at the baseline, which are still considered as non PFOA cases, might experience worsening of an existing abnormality in the following years and diagnosed with PFOA (progression).
Similarly, knees that did not show any signs of PFOA at the baseline might develop the disease for the first time during the the following 7 years (incidence).
In MOST, individual radiographic features were  graded by two independent expert readers and when there was a disagreement in film readings, a panel of three adjudicators resolved the discrepancies \cite{roemer2009association}.

%TC:ignore
\begin{figure}[!t]
\centerline{
\begin{adjustbox}{color = blue, varwidth=1.1\textwidth,margin=3pt 1pt 3pt 5,frame=0.5pt }
\centering
\includegraphics[height=5.3cm]{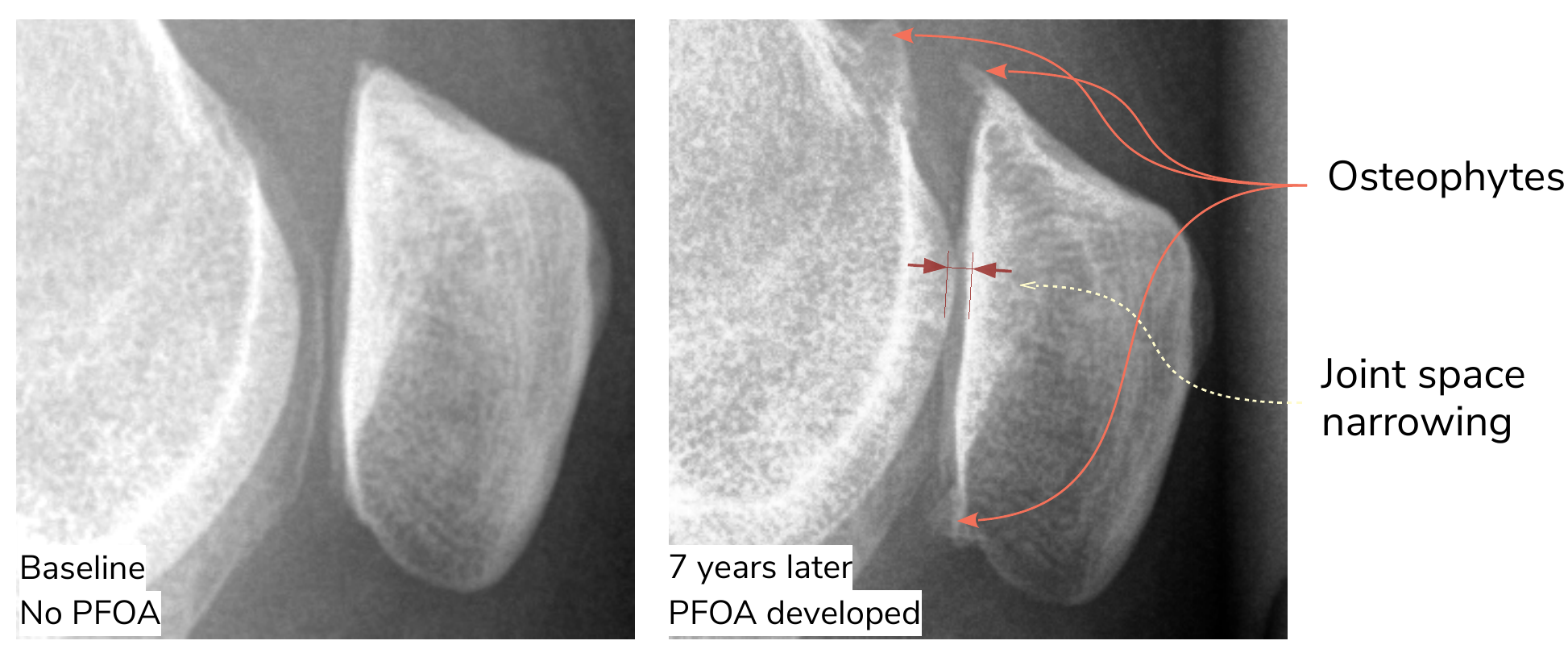}

\caption{ Example of PFOA progression/development.
Figure on the left demonstrates an exemplar patellofemoral joint ROI imaged at the first visit in the MOST study. At the baseline, PFOA is not present. Right figure presents the same participant's PF joint 7 years after the baseline visit. 
The knee has developed PFOA where joint  space  narrowing  (JSN) and osteophytes - characteristic features of OA - are clearly seen.
Best viewed on screen.
}
\label{fig:sample}

\end{adjustbox}
}
\end{figure}
%TC:endignore

\subsection{Selection of Region of Interest (ROI)}
\begin{figure}[!t]
\centerline{
\begin{adjustbox}{color = blue, varwidth=1.1\textwidth,margin=3pt 1pt 3pt 5,frame=0.5pt }
\centering
\includegraphics[height=6.3cm]{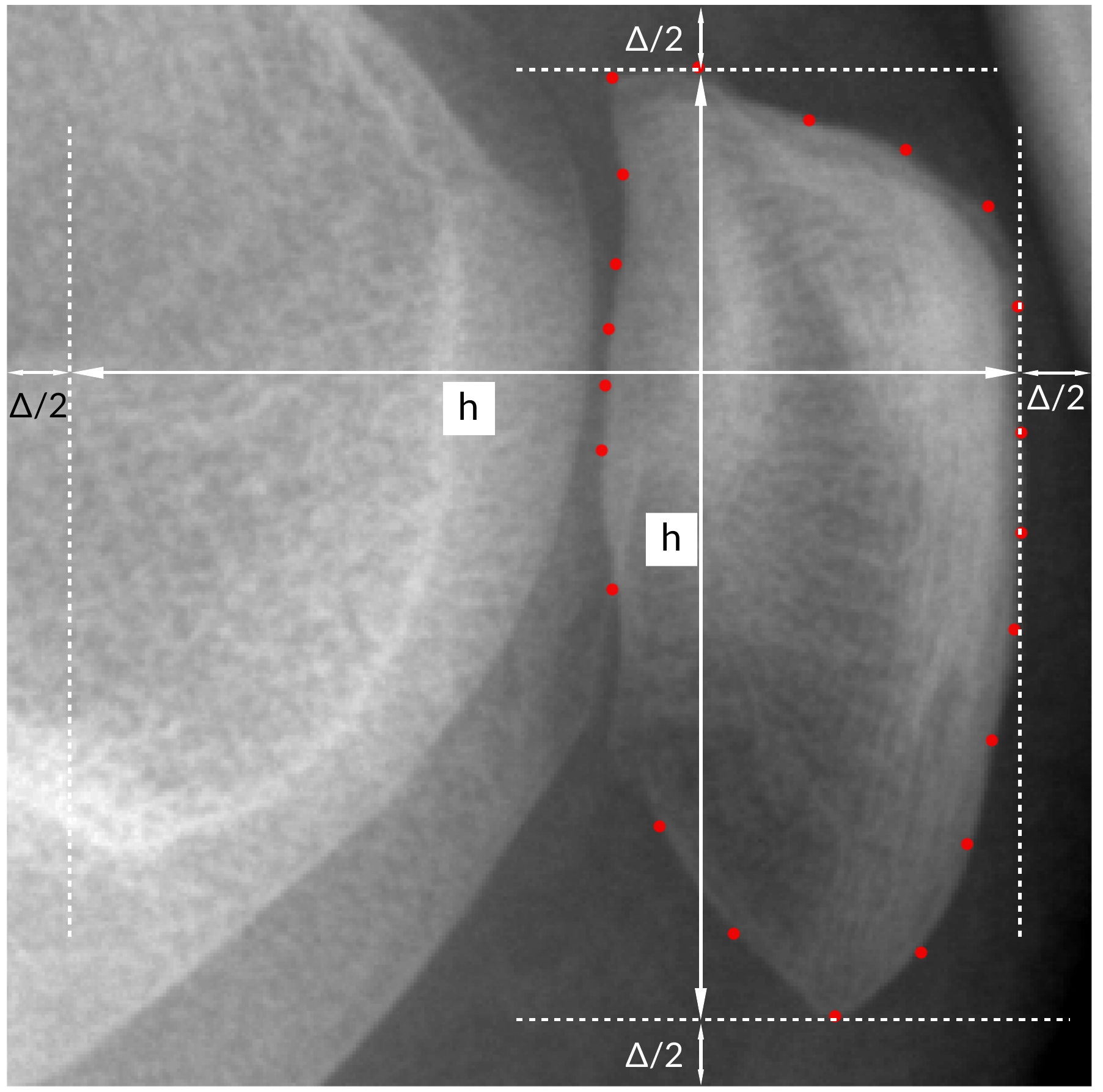}

\caption{ Illustration of automated ROI localization. 
First, patellar height (h) was determined using landmarks.
Subsequently, a small margin ($\Delta$) is padded around the patellar region.
On the femur side, ROI is located such that the width equals to the height of the ROI.
Best viewed on screen.
}
\label{fig:roi}

\end{adjustbox}
}
\end{figure}

We placed a PFJROI automatically using landmarks (Figure \ref{fig:roi}).
The height of the patellar bone ($h$) was used to locate a square shaped image ROI.
Once the patellar bone margins were determined using landmarks, a $20$pixels ($\Delta\over2$) region is padded around the bone. 
On the femur side, the ROI is extended to capture the part of the femur facing the patellar bone such that the width of the ROI equals to the height of it ($height= width= h+\Delta$).
Finally, the size of the ROI becomes proportional to the size of the patellar bone.
%For details of PFJROI see Supplementary.

\subsection{Predicting progression of patellofemoral osteoarthritis using Deep CNN}
We adopted the deep CNN architecture proposed by Yan et al. \cite{yan2019melanoma} to predict PFOA development based on the baseline imaging data.
It uses VGG-16 \cite{simonyan2014very} backbone with  two additional attention layers and one penultimate global feature
vector (obtained via global average pooling)(Figure \ref{fig:flowchart}). PFJROI data were pre-processed by resizing it to $256\times256$ pixels and then applying a random crop of size $224\times224$ pixels.
The backbone network VGG-16 was initialized with its pre-trained version on ImageNet.
The attention modules were initialized using He's initialization \cite{he2015delving}.
We employed Focal loss \cite{lin2017focal}, a variant of the cross-entropy, which has shown to be an effective when facing the class imbalance problem  by selectively downweighting well-classified examples.
We used a batch size of 32 and trained the network end-to-end for 45 epochs
using stochastic gradient descent with momentum. 
The initial learning rate was 0.001 and it was decayed by 10 every 10 epochs. 

To examine the impact of the attention mechanism on the model's performance, a separate training was conducted with the original VGG-16 network without the attention modules. The network parameters were initialized with ImageNet pre-training, and the last layer was modified for binary classification. To ensure a fair comparison, we maintained consistency in the other network parameters and hyper-parameters between the attention model and the model without attention.

\subsection{Attention Module}
Previous deep learning works that employ post hoc analysis for visual explanations such as Grad-CAM \cite{selvaraju2017grad} require extra computation based on a fully trained classification network and relies on  gradient information passed to the last convolutional layer combined with the forward activation maps.
However, those feature maps, that are often used to produce explanations, are not necessarily related to the target class and they do not affect the network parameters at all.
In this study, we employed a trainable spatial attention mechanism to produce insights into the model decisions.
Attention mechanisms are widely used in the field of natural language processing (NLP) as a way to improve the performances of models by emphasizing the important parts of the information \cite{vaswani2017attention}.
In case of image classification, the idea of trainable attention is to focus on the most informative parts of an image while ignoring less relevant or noisy parts.
During training, the network learns to weight different regions of the input image based on the classification performance.
See Supplementary Material for more details of the attention module used in our architecture.

\subsection{Reference Models}
We employed GBM to predict the development of PFOA from demographic data and self-reported symptom assessments.
GBM is a popular and powerful machine learning algorithm used for regression and classification based on ensembles of decision trees\cite{friedman2001greedy}.
It works iteratively by adding decision trees to the model where each new tree attempts to correct the errors made by the previous trees.  
In this study, we used an efficient implementation of GBM called LightGBM \cite{ke2017lightgbm}.

We built three GBM classifiers based on the clinical data and risk factors.
These include age, sex, body-mass index (BMI), the total Western Ontario and McMaster Universities Arthritis Index (WOMAC) score, and the KL grade of the tibiofemoral joint  (Model1, Model2, and Model3 in Table \ref{tab:comparison}).
The WOMAC score is a widely used questionnaire-based assessment tool designed to evaluate the severity of pain, stiffness, and physical disability in patients with OA of the knee and hip.

For all of our models, we utilized subject-wise stratified 5-fold cross validation. This involves dividing the dataset into 5 folds, each containing data from different subjects, and stratifying the data within each fold so that the proportion of progressors vs non-progressors is similar to the overall dataset. This helps to eliminate subject-dependent bias between the training and validation sets.

K-fold cross-validation involves iteratively selecting one fold as the testing set and the remaining folds as the training set. The model is trained on the training set and evaluated on the testing set. This process is repeated for each fold, with each fold serving as the testing set exactly once.

To ensure fair comparisons, we used the same folds for all of the models. All of the models were trained separately and the reported performances were derived from these separate models.

\subsection{Statistical Methods}

The performance of the models were compared using Receiver Operating Characteristics (ROC) curves, Precision-Recall (PR) curves, and Brier score \cite{brier1950verification}.
ROC curves plot the true positive rate (TPR) against the false positive rate (FPR) at various classification thresholds. The area under the ROC curve (AUC-ROC) is often used as a summary metric for model performance, with a value of 1 indicating perfect classification and 0.5 indicating random classification.
On the other hand, PR curves plot the precision (positive predictive value) against the recall (true positive rate) at various classification thresholds. The area under the PR curve (average precision, AP) is another commonly used summary metric for model performance, with a value of 1 indicating perfect classification and 0 indicating random classification.
ROC curves are often used when the number of negative instances is much larger than the number of positive instances, while PR curves are more suitable when the number of positive instances is relatively small. In general, a good classifier should have high values for both AUC-ROC and AUC-PR.
Brier score equals to the mean squared error of the prediction.
In order to compare the differences between model AUCs, we applied DeLong's test \cite{delong1988comparing}.

\section{Results}

\begin{figure} [!htb]
    \centerline{
    % \hspace{-1cm}
    \includegraphics[width=0.7\textwidth]{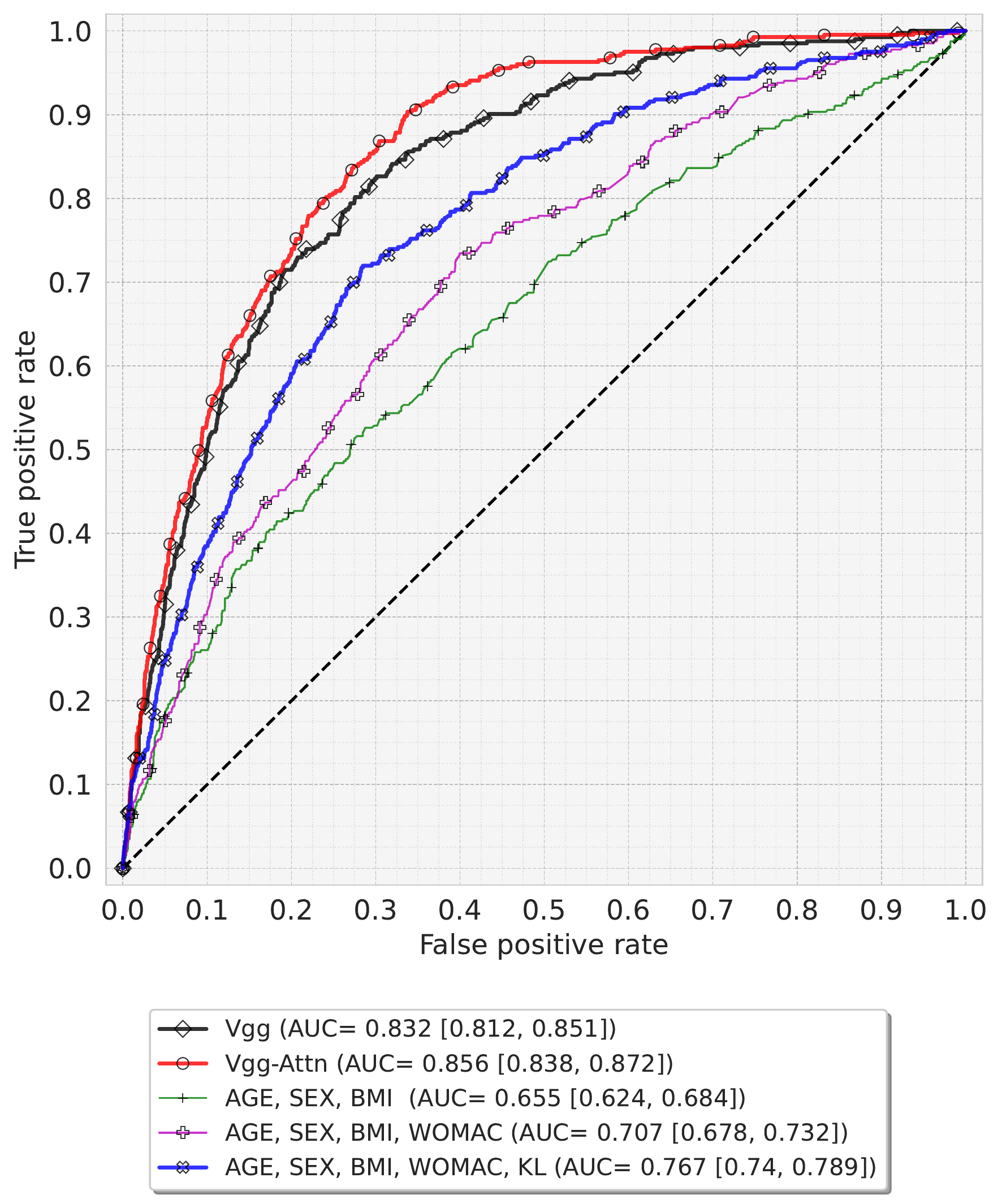}
    \includegraphics[width=0.7\textwidth]{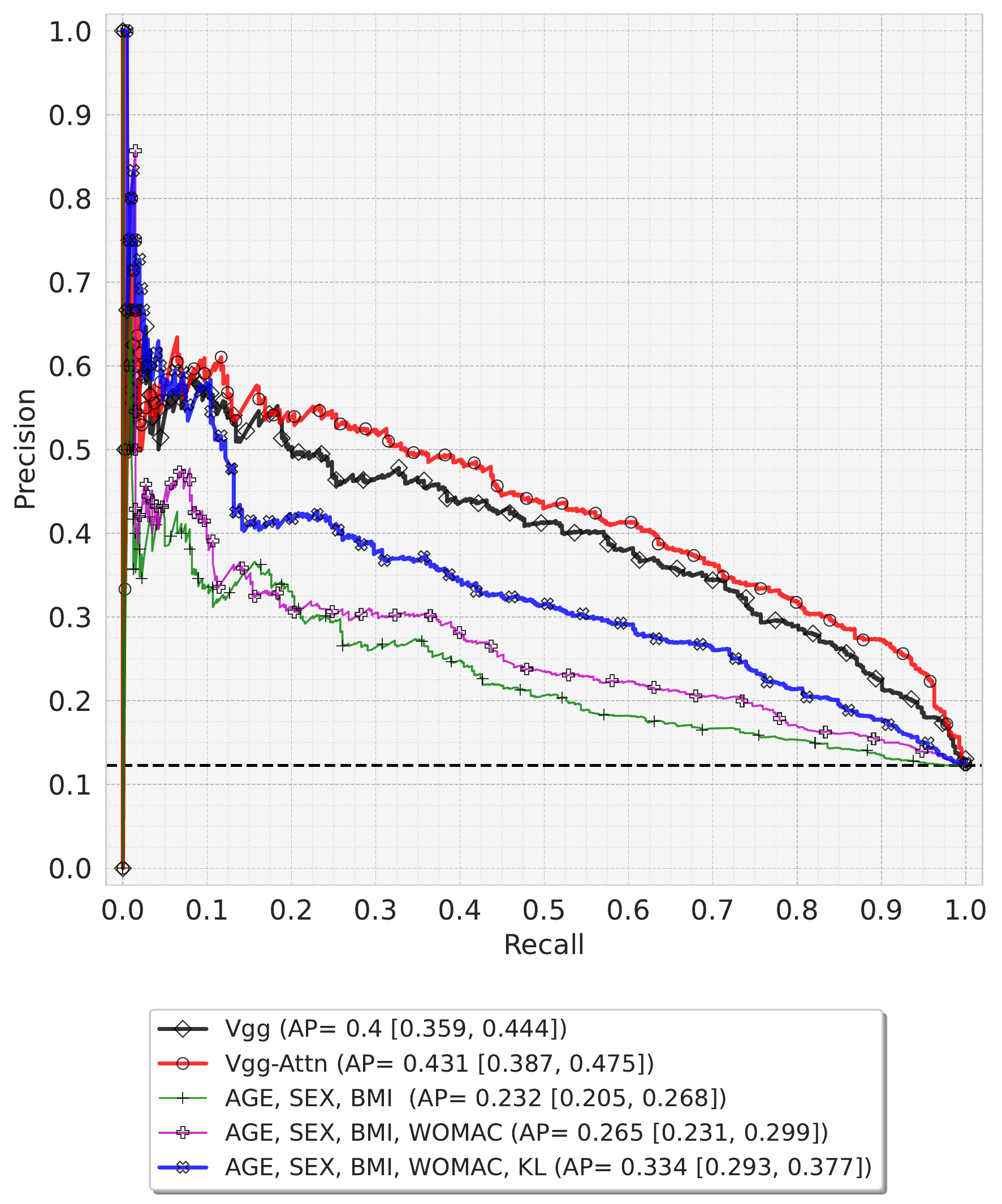}\hfill}
    \caption{ (a) ROC and (b) PR curves demonstrating the performance of the models. 
    Area under the curves and 95\% confidence intervals in parentheses were given based on a 5-fold cross validation setting.
    Dashed lines in ROC indicate the performance of a random classifier and in case of PR it indicates the distributions of the labels of the dataset (Progressor vs non-progressor).}
    \label{fig:perf}
\end{figure}

Table \ref{tab:comparison} and Figure \ref{fig:perf} show the performance of different models in predicting PFOA progression. Our proposed VGG-16-Attn model achieved the highest AUC of 0.856 [0.838, 0.872] and AP of 0.431 [0.387, 0.475] among all the considered models (Model1 to Model5). We compared the performance of VGG-16-Attn with the original VGG-16 model to assess the contribution of attention modules. Our results show that the addition of attention modules has a positive impact on the performance of the model, with a statistically significant difference between the AUC values of the two models (DeLong's p-value $=0.00018$).

To assess the value of imaging biomarkers in predicting PFOA progression, we conducted a thorough evaluation of various risk factors, including age, sex, body-mass index (BMI), WOMAC, and TFOA KL scores (Figure \ref{fig:perf}) as reference models. Using gradient boosting machine (GBM) models, we trained the models to predict the probability of developing PFOA based on different combinations of these risk factors. Our results showed that the best-performing reference model (Model3) incorporated age, sex, BMI, WOMAC, and TFOA KL scores, achieving an AUC of 0.767 [0.74, 0.789] and an AP of 0.334 [0.293, 0.377] (Figure \ref{fig:perf}).
We also measured the impact of each feature on the model's output by looking at the contribution of that feature to the predicted outcome compared to what the predicted outcome would be if the feature was not included in the model (SHapley Additive exPlanations \cite{NIPS2017} (Supplementary Figure \ref{fig:fi} and Supplementary Figure \ref{fig:sum}). High BMI, WOMAC and KL scores increase the predicted PFOA progression risk and low BMI, WOMAC, KL scores reduce the risk.

Subsequently, we compared the performance of our deep convolutional neural network (CNN) attention model (VGG-16-Attn, Model5) to the best-performing reference method (Model3). Our results showed a statistically significant difference between the AUC values of the two models (DeLong's p-value$<1e-10$).

To further improve predictive accuracy, we used a second-layer GBM model that fused the predictions of the VGG-16-Attn CNN model (Model5) and the strongest reference model (Model3) with imaging features and clinical assessments (Figure \ref{fig:flowchart}c). This stacked model (Model6) achieved the best AUC of 0.865 [0.838, 0.872], an AP of 0.447 [0.404, 0.491], and a Brier score of 0.084, outperforming both individual models. While the increase in AUC between the stacked model (Model6) and the VGG-16-Attn CNN model (Model5) was statistically significant (DeLong's p-value $=0.0085$), it was not highly significant.

\begin{table}[]
\caption{Comparison of the developed models.   AUC and  AP  indicate  the  area  under  the  Receiver  Operating  Characteristics  (ROC)  curve  and the area under the Precision-Recall (PR) curves, respectively.
The 95\% confidence intervals in parentheses were given based on a 5-fold cross validation setting.}
\label{tab:comparison}
\centerline{
\resizebox{1.2\columnwidth}{!}{%
\begin{tabular}{@{}lllllll@{}}
\toprule
                            & Input                                                                                                               & Method                   & AUC [95\% CI]                                          & AP [95\% CI]                                            & \begin{tabular}[c]{@{}l@{}}Brier \\ Score\end{tabular} &                \\ \midrule
\multicolumn{1}{l|}{Model1} & \multicolumn{1}{l|}{Age, Sex, BMI}                                                                                  & \multicolumn{1}{l|}{GBM} & \multicolumn{1}{l|}{0.655 {[}0.624, 0.684{]}} & \multicolumn{1}{l|}{0.232 {[}0.205, 0.268{]}} & \multicolumn{1}{l|}{0.103}                             & Clinical Model \\ \midrule
\multicolumn{1}{l|}{Model2} & \multicolumn{1}{l|}{Age, Sex, BMI, WOMAC}                                                                           & \multicolumn{1}{l|}{GBM} & \multicolumn{1}{l|}{0.707 {[}0.678, 0.732{]}} & \multicolumn{1}{l|}{0.265 {[}0.231, 0.299{]}} & \multicolumn{1}{l|}{0.100}                             & Clinical Model \\ \midrule
\multicolumn{1}{l|}{Model3} & \multicolumn{1}{l|}{Age, Sex, BMI, WOMAC, KL}                                                                       & \multicolumn{1}{l|}{GBM} & \multicolumn{1}{l|}{0.767 {[}0.74, 0.789{]}} & \multicolumn{1}{l|}{0.334 {[}0.293, 0.377{]}} & \multicolumn{1}{l|}{0.095}                             & Clinical Model \\ \midrule

\multicolumn{1}{l|}{Model4} & \multicolumn{1}{l|}{\begin{tabular}[c]{@{}l@{}}VGG-16\end{tabular}}                           & \multicolumn{1}{l|}{CNN} & \multicolumn{1}{l|}{0.832 {[}0.812, 0.851}      & \multicolumn{1}{l|}{0.4 {[}0.359, 0.444{]}} & \multicolumn{1}{l|}{0.262}                             & CNN model      \\ \midrule

\multicolumn{1}{l|}{Model5}                     & \multicolumn{1}{l|} {\begin{tabular}[c]{@{}l@{}} VGG-16-Attn\end{tabular} }                                      &  \multicolumn{1}{l|} {CNN}                      & \multicolumn{1}{l|}{{0.856 [0.838, 0.872]}   }          & \multicolumn{1}{l|}{{0.431 [0.387, 0.475]} }            & \multicolumn{1}{l|}{0.165}                                         & CNN Model  \\ \midrule

\multicolumn{1}{l|}{Model6}                     & \multicolumn{1}{l|} {\begin{tabular}[c]{@{}l@{}}Predictions from \\ Model3 and Model5\end{tabular} }                                      &  \multicolumn{1}{l|} {GBM}                      & \multicolumn{1}{l|}{\textbf{0.865 {[}0.849, 0.88{]}}   }          & \multicolumn{1}{l|}{\textbf{0.447 {[}0.404, 0.491{]}} }            & \multicolumn{1}{l|}{\textbf{0.084}}                                         & Stacked Model  \\ \bottomrule
\end{tabular}
}
}
\end{table}

\begin{figure}[!htb]
    \centerline{
    % \hspace{-1cm}
    \includegraphics[width=0.7\textwidth]{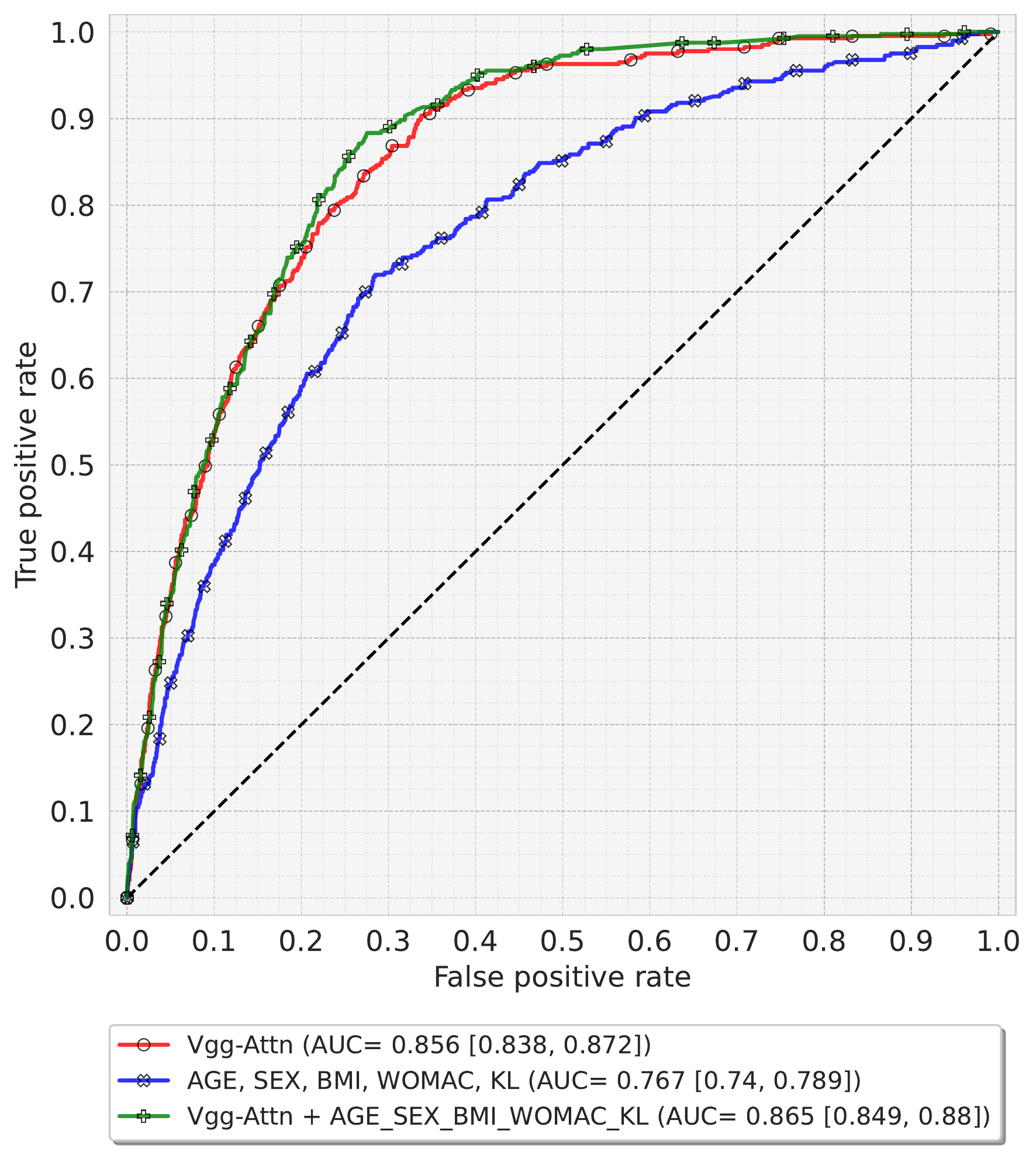}
    \includegraphics[width=0.7\textwidth]{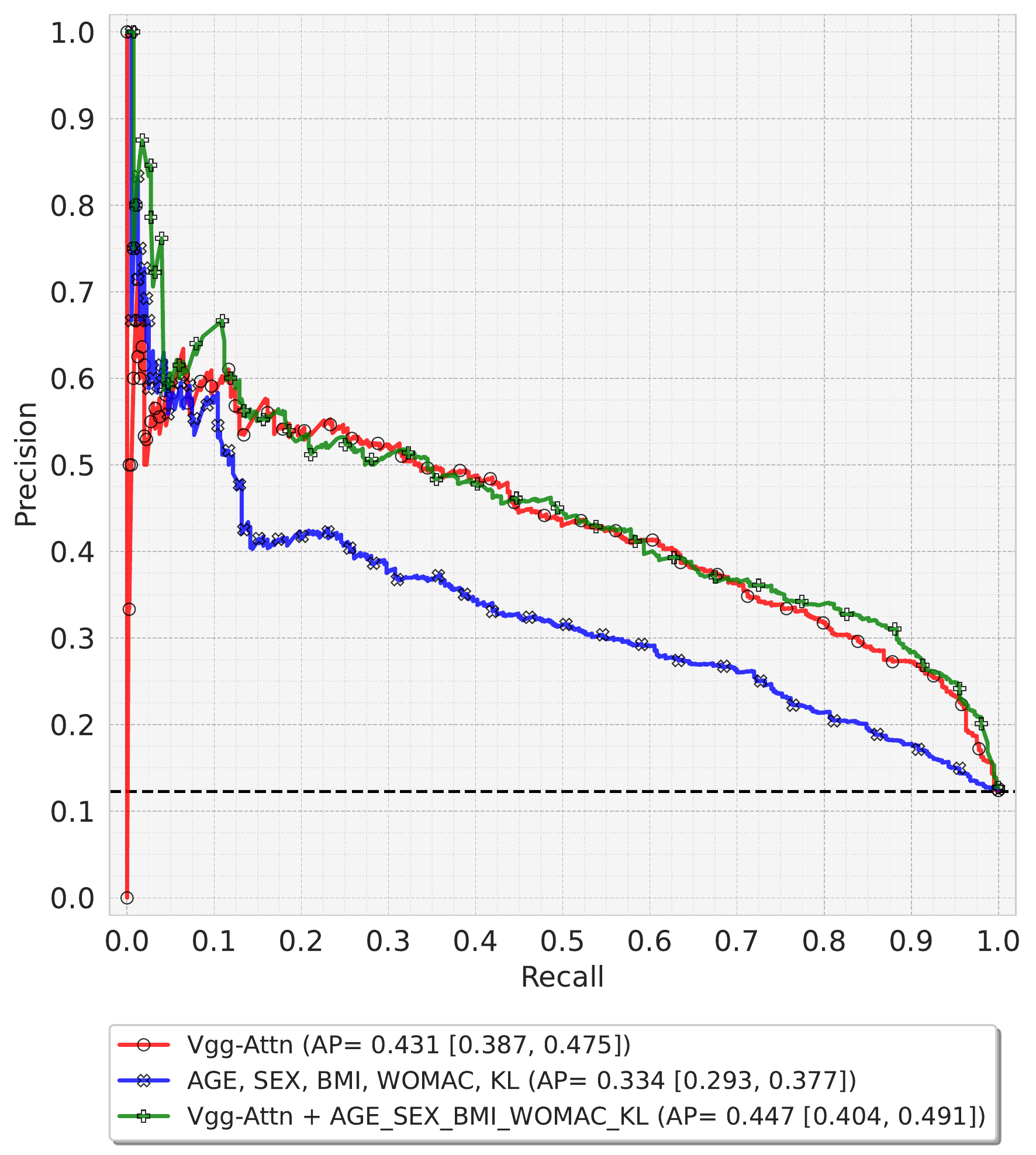}
    \hfill}
    \caption{ (a) ROC and (b) PR curves demonstrating the performance of the attention model (VGG-16-Attn), best clinical model (Model3) ) and stacked model (Model6). 
    Area under the curves and 95\% confidence intervals in parentheses were given based on a 5-fold cross validation setting.
    Dashed lines in ROC indicate the performance of a random classifier and in case of PR it indicates the distributions of the labels of the dataset (PFOA vs non-PFOA).}
    \label{fig:stacked}
\end{figure}

\begin{figure}[!htb]
    \centerline{
    % \hspace{-1cm}
    \includegraphics[width=1\textwidth]{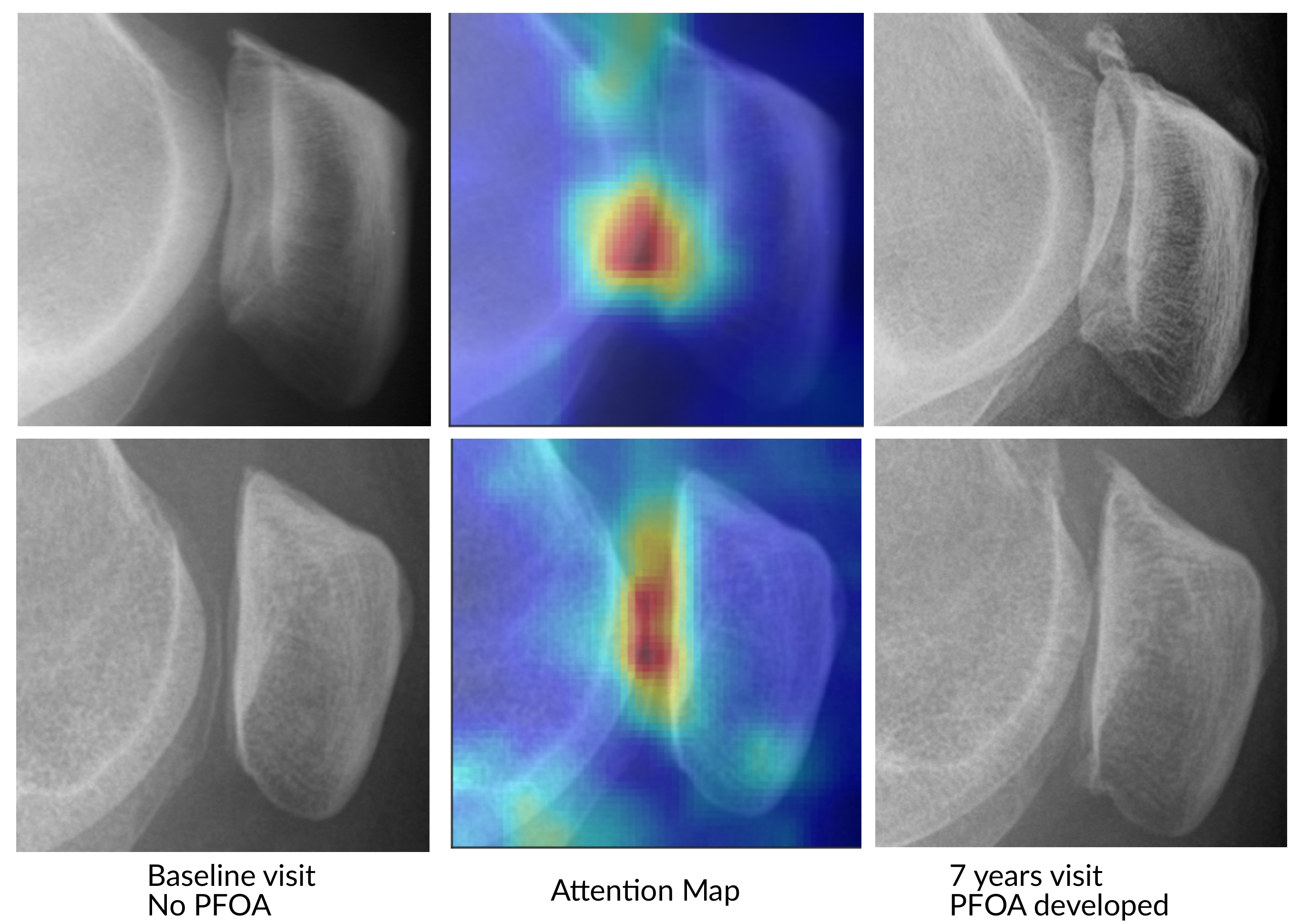}
    \hfill}
    \caption{ Examples of attention maps of the two progressor knees from the dataset. First  column shows the baseline radiographs in which the knee does not have PFOA yet. Middle column illustrates the attention maps and finally last column presents the final follow-up radiographs.}
    \label{fig:attention_examples}
\end{figure}

Examples of spatial attention maps are presented in Figure \ref{fig:attention_examples}. 
The shallower attention map which is applied after conv3 layer, focus on more general and diffused areas.
Therefore, we present here only the deeper attention map (after the conv4 layer in Figure \ref{fig:flowchart}).
In various cases, the model paid attention to the PF joint space width and the inferior and posterior regions of patellar bone.
Additional examples of such attention maps are presented in the Supplementary. 

\section{Discussion}

This study presents a novel deep learning-based approach for predicting progression of PFOA, utilizing both clinical variables and imaging data. 
The results demonstrate the potential of machine learning techniques, especially deep learning, in predicting PFOA progression, which could provide valuable information for clinicians in patient care.

In general, ML-based models can handle heterogeneous data and they can identify patterns that may not be apparent to human experts.
We highlighted this by the inclusion of both clinical variables and imaging data into the stacked model.
This combination model achieved the highest accuracy in predicting PFOA progression, indicating its ability to differentiate between patients who are likely to experience PFOA and those who are not. However, it should be still noted that the performance gain with the stacked model (AUC=0.865, AP=0.447), compared to the imaging-based model (AUC=0.856, AP=0.431), was only minor and, although statistically significant, probably the clinical gain might be insignificant. Consequently, this suggests that clinical variables have only minor contribution to the prediction performance on top of the X-ray image alone. Similarly as in the case of knee OA progression prediction\cite{tiulpin2019multimodal}, it looks like that a knee lateral X-ray image already includes indirectly a lot of clinical information, such as age and BMI.

\begin{figure}[hbt!]
\centering
\subfloat[Model1]{\includegraphics[width = 0.8\linewidth, ]{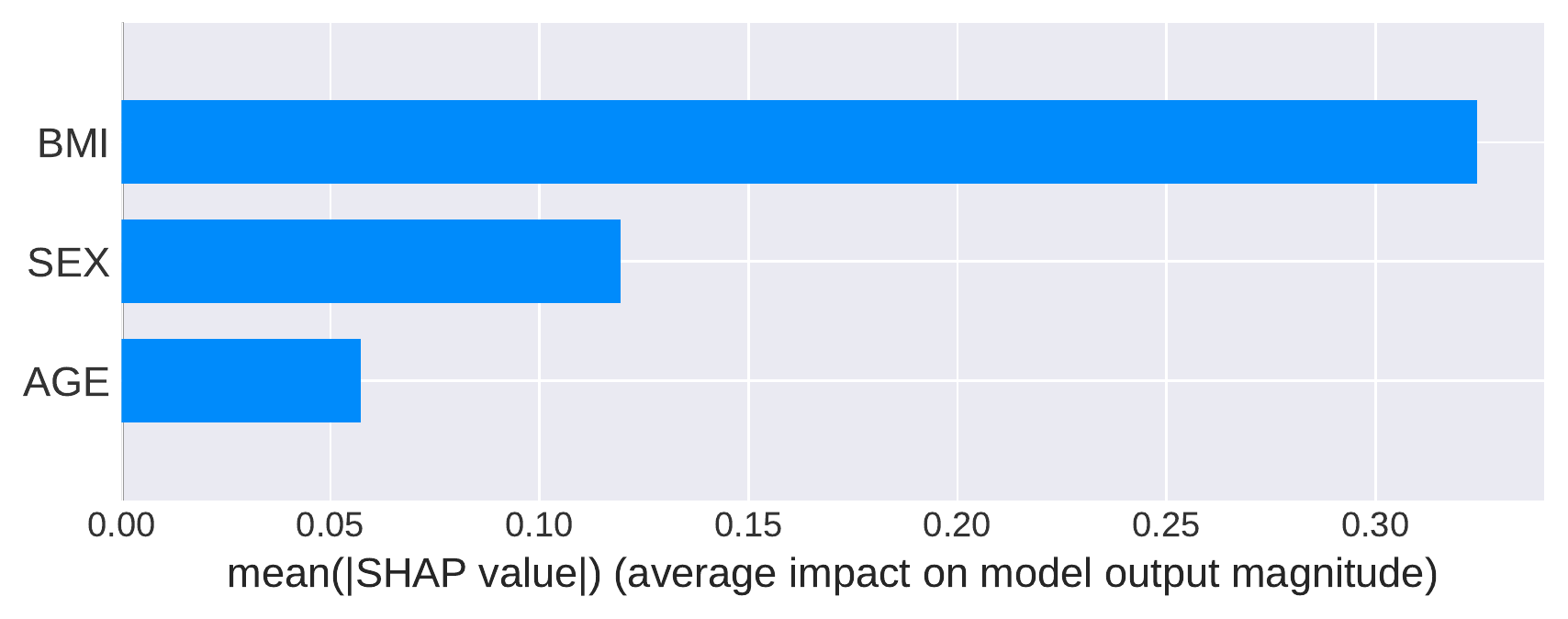}}\\
\subfloat[Model2]{\includegraphics[width = 0.8\linewidth, ]{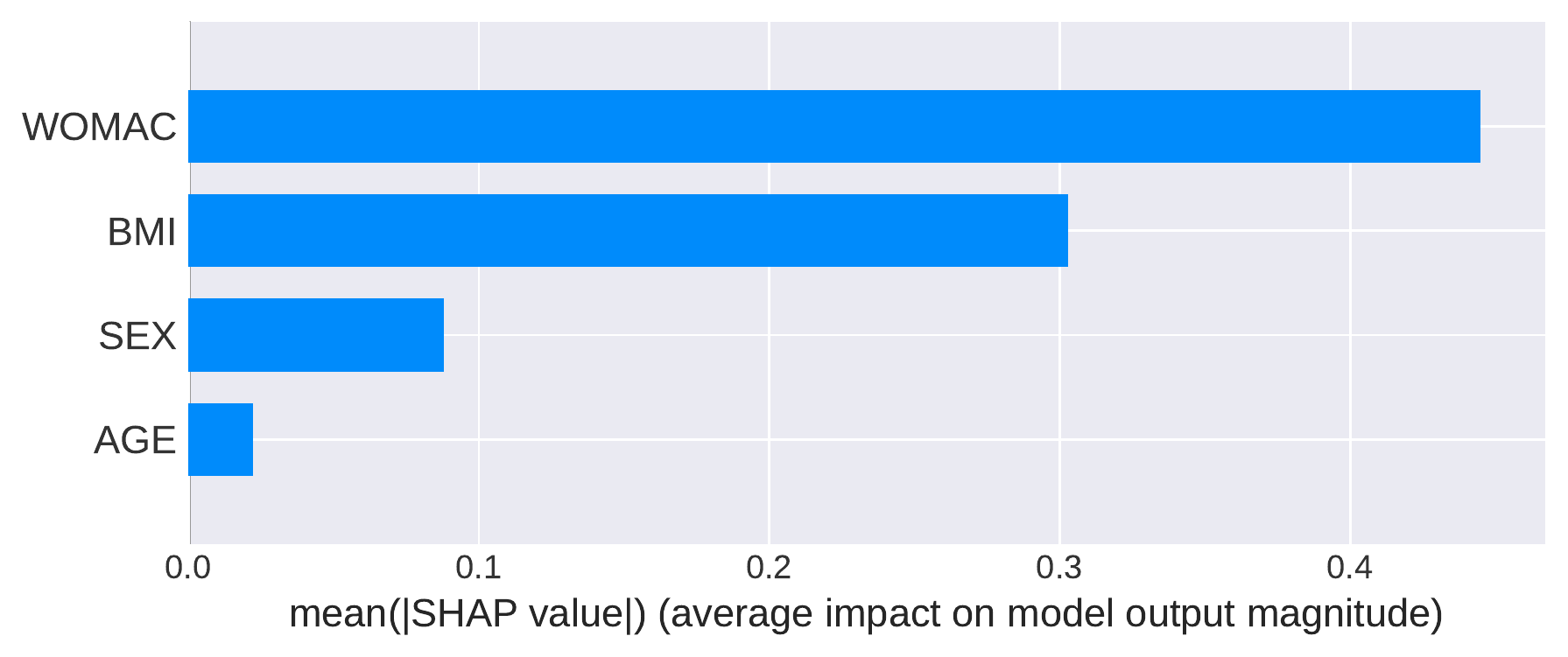}}\\
\subfloat[Model3]{\includegraphics[width = 0.8\linewidth, ]{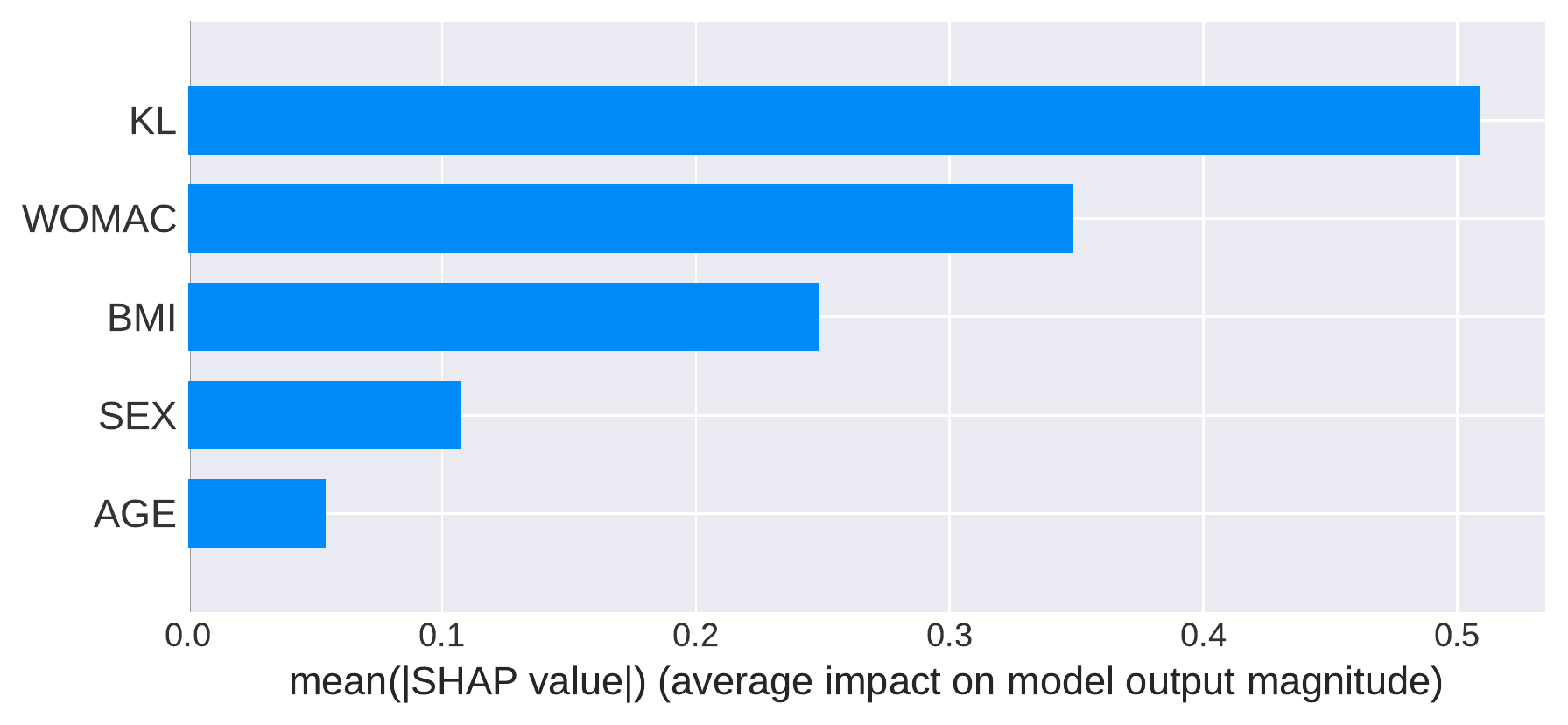}}\\
\caption{SHAP feature importance plots. It is measured as the mean absolute Shapley values. a) BMI in Model1 , b) WOMAC score in Model2 and c) Tibiofemoral joint OA status (KL score) in Model3 is the most important feature.}
\label{fig:fi}
\end{figure}

Our study confirmed that high BMI, high WOMAC score, female sex, and OA in the TF joint (KL score) are all risk factors for PFOA development (Supplementary Figures \ref{fig:fi} and \ref{fig:sum} and (Table \ref{tab:demographics}). 
Out of the three main demographical variables age, sex and BMI in isolation (Model1), the strongest predictive capability was high BMI.

It has been earlier reported that the use of attention mechanism increases the performance of NLP models \cite{niu2021review, vaswani2017attention}. Here, we also observed the increased performance in this kind of image classification task (AUC = 0.856 vs. 0.832, AP = 0.431 vs. 0.400). Besides the increase in overall model performance, generated attention maps highlighted the joint space and the regions where osteophytes typically occur. These regions are known to be affected in PFOA, and they reflect manual imaging biomarkers of OA including joint space narrowing and morphological and structural changes in bone.

The present study is unique as it investigated the potential of machine learning approaches based on imaging data to accurately predict PFOA progression for the first time. However, there are also some limitations of this study. First and foremost, the model was trained on data from a single population, and further research is necessary to validate the model's generalizability to other populations and settings. Additionally, the study did not consider other potential predictors of PFOA progression, such as biomechanical or genetic factors. Incorporating longitudinal data and other types of imaging data, such as MRI, could further improve the model.

In conclusion, our study demonstrates the potential of machine learning models to predict PFOA progression using imaging and clinical variables. These models could assist in identifying patients at high risk of PFOA progression, enabling clinicians to intervene with personalized treatment plans and potentially prevent or delay disease progression.

\section*{Summary Table}
\begin{itemize}
\item We present the first study for predicting PFOA progression based on a multi-modal machine learning method using lateral X-ray images and clinical data.
\item We leveraged trainable attention mechanism to highlight regions in lateral X-rays which highly contributed the decision of the model.

\item We compared the performances of deep convolutional neural network based models and gradient boosting machine based models using
clinical variables including age, sex, body mass index (BMI), the total Western Ontario and McMaster Universities ArthritisIndex (WOMAC) score, and Kellgren and Lawrence (KL) score of the tibiofemoral joint.

\item Finally,  we  proposed  a  stacked  model  where  both deep CNN predictions and predictions from clinical model are combined with a second level machine learning model - Gradient Boosting Machine (GBM).

\item Our  results  demonstrated that imaging biomarkers contain useful information for predicting PFOA progression within 7 years. Moreover, addition of clinical data slightly improves the prediction power of the imaging-based model, although the clinical significance of this performance gain is unknown.  

\item Predicting PFOA progression/ development have the potential to improve the early diagnosis, management and treatment of PFOA.

\end{itemize}

\section*{Acknowledgements}%

Multicenter Osteoarthritis Study (MOST) Funding Acknowledgment. MOST is comprised of four cooperative grants (Felson – AG18820; Torner – AG18832, Lewis – AG18947, and Nevitt – AG19069) funded by the National Institutes of Health, a branch
of the Department of Health and Human Services, and conducted by MOST study investigators. This manuscript was prepared using MOST data and does not necessarily reflect the opinions or views of MOST investigators.

We would like to acknowledge the NORDFORSK grant from the project ``Molecular and structural biomarkers for personalised care in osteoarthritis" (Project No.: 116406).

We gratefully acknowledge Claudia Lindner for providing the BoneFinder\textsuperscript{\textregistered} tool and lateral knee active shape model, Aleksei Tiulpin for providing an interface to BoneFinder to fully leverage multiple processors, and the support of NVIDIA Corporation with the donation of the Quadro P6000 GPU used in this research.

\section*{Funding}%
Funding sources are not associated with the scientific contents of the study.

\section*{Competing interests}
The authors declare that they have no competing interests.

\section*{Authors' contributions}
 N.B. originated the idea of the study. N.B. performed
the experiments and took major part in writing of the manuscript. S.S. supervised the project. All  authors participated in producing the final manuscript draft and approved the final submitted version.

\bibliography{bib.bib}

\begin{thebibliography}{10}
\expandafter\ifx\csname url\endcsname\relax
  \def\url#1{\texttt{#1}}\fi
\expandafter\ifx\csname urlprefix\endcsname\relax\def\urlprefix{URL }\fi
\expandafter\ifx\csname href\endcsname\relax
  \def\href#1#2{#2} \def\path#1{#1}\fi

\bibitem{lankhorst2017incidence}
N.~Lankhorst, J.~Damen, E.~Oei, J.~Verhaar, M.~Kloppenburg, S.~Bierma-Zeinstra,
  M.~van Middelkoop, Incidence, prevalence, natural course and prognosis of
  patellofemoral osteoarthritis: the cohort hip and cohort knee study,
  Osteoarthritis and Cartilage 25~(5) (2017) 647--653.

\bibitem{duncan2008pain}
R.~Duncan, G.~Peat, E.~Thomas, L.~Wood, E.~Hay, P.~Croft, How do pain and
  function vary with compartmental distribution and severity of radiographic
  knee osteoarthritis?, Rheumatology 47~(11) (2008) 1704--1707.

\bibitem{crossley2011patellofemoral}
K.~Crossley, R.~Hinman, The patellofemoral joint: the forgotten joint in knee
  osteoarthritis, Osteoarthritis and Cartilage 19~(7) (2011) 765--767.

\bibitem{duncan2011incidence}
R.~Duncan, G.~Peat, E.~Thomas, E.~Hay, P.~Croft, Incidence, progression and
  sequence of development of radiographic knee osteoarthritis in a symptomatic
  population, Annals of the rheumatic diseases 70~(11) (2011) 1944--1948.

\bibitem{duncan2009does}
R.~Duncan, G.~Peat, E.~Thomas, L.~Wood, E.~Hay, P.~Croft, Does isolated
  patellofemoral osteoarthritis matter?, Osteoarthritis and cartilage 17~(9)
  (2009) 1151--1155.

\bibitem{kim2012patellofemoral}
Y.-M. Kim, Y.-B. Joo, Patellofemoral osteoarthritis, Knee Surgery and Related
  Research 24~(4) (2012) 193--200.

\bibitem{van2018international}
M.~van Middelkoop, K.~L. Bennell, M.~J. Callaghan, N.~J. Collins, P.~G.
  Conaghan, K.~M. Crossley, J.~J. Eijkenboom, R.~A. van~der Heijden, R.~S.
  Hinman, D.~J. Hunter, et~al., International patellofemoral osteoarthritis
  consortium: consensus statement on the diagnosis, burden, outcome measures,
  prognosis, risk factors and treatment, in: Seminars in Arthritis and
  Rheumatism, Vol.~47, Elsevier, 2018, pp. 666--675.

\bibitem{kobayashi2016prevalence}
S.~Kobayashi, E.~Pappas, M.~Fransen, K.~Refshauge, M.~Simic, The prevalence of
  patellofemoral osteoarthritis: a systematic review and meta-analysis,
  Osteoarthritis and cartilage 24~(10) (2016) 1697--1707.

\bibitem{macri2017patellofemoral}
E.~M. Macri, Patellofemoral osteoarthritis: characterizing knee alignment and
  morphology, Ph.D. thesis, University of British Columbia (2017).

\bibitem{peat2012clinical}
G.~Peat, R.~C. Duncan, L.~R. Wood, E.~Thomas, S.~Muller, Clinical features of
  symptomatic patellofemoral joint osteoarthritis, Arthritis research \&
  therapy 14~(2) (2012) R63.

\bibitem{de2016radiographic}
B.~de~Lange-Brokaar, J.~Bijsterbosch, P.~Kornaat, E.~Yusuf, A.~Ioan-Facsinay,
  A.-M. Zuurmond, H.~Kroon, I.~Meulenbelt, J.~Bloem, M.~Kloppenburg,
  Radiographic progression of knee osteoarthritis is associated with mri
  abnormalities in both the patellofemoral and tibiofemoral joint,
  Osteoarthritis and cartilage 24~(3) (2016) 473--479.

\bibitem{bayramoglu2021pfoa}
N.~Bayramoglu, M.~T. Nieminen, S.~Saarakkala, Automated detection of
  patellofemoral osteoarthritis from knee lateral view radiographs using deep
  learning: data from the multicenter osteoarthritis study (most),
  Osteoarthritis and Cartilage 29~(10) (2021) 1432--1447.

\bibitem{bayramoglu2022machine}
N.~Bayramoglu, M.~T. Nieminen, S.~Saarakkala, Machine learning based texture
  analysis of patella from x-rays for detecting patellofemoral osteoarthritis,
  International journal of medical informatics 157 (2022) 104627.

\bibitem{lindner2013fully}
C.~Lindner, S.~Thiagarajah, J.~M. Wilkinson, G.~A. Wallis, T.~F. Cootes,
  arcOGEN Consortium, et~al., Fully automatic segmentation of the proximal
  femur using random forest regression voting, IEEE transactions on medical
  imaging 32~(8) (2013) 1462--1472.

\bibitem{friedman2001greedy}
J.~H. Friedman, Greedy function approximation: a gradient boosting machine,
  Annals of statistics (2001) 1189--1232.

\bibitem{roemer2009association}
F.~Roemer, A.~Guermazi, D.~Hunter, J.~Niu, Y.~Zhang, M.~Englund, M.~Javaid,
  J.~Lynch, A.~Mohr, J.~Torner, et~al., The association of meniscal damage with
  joint effusion in persons without radiographic osteoarthritis: the framingham
  and most osteoarthritis studies, Osteoarthritis and cartilage 17~(6) (2009)
  748--753.

\bibitem{yan2019melanoma}
Y.~Yan, J.~Kawahara, G.~Hamarneh, Melanoma recognition via visual attention,
  in: Information Processing in Medical Imaging: 26th International Conference,
  IPMI 2019, Hong Kong, China, June 2--7, 2019, Proceedings 26, Springer, 2019,
  pp. 793--804.

\bibitem{simonyan2014very}
K.~Simonyan, A.~Zisserman, Very deep convolutional networks for large-scale
  image recognition, arXiv preprint arXiv:1409.1556.

\bibitem{he2015delving}
K.~He, X.~Zhang, S.~Ren, J.~Sun, Delving deep into rectifiers: Surpassing
  human-level performance on imagenet classification, in: Proceedings of the
  IEEE international conference on computer vision, 2015, pp. 1026--1034.

\bibitem{lin2017focal}
T.-Y. Lin, P.~Goyal, R.~Girshick, K.~He, P.~Doll{\'a}r, Focal loss for dense
  object detection, in: Proceedings of the IEEE international conference on
  computer vision, 2017, pp. 2980--2988.

\bibitem{selvaraju2017grad}
R.~R. Selvaraju, M.~Cogswell, A.~Das, R.~Vedantam, D.~Parikh, D.~Batra,
  Grad-cam: Visual explanations from deep networks via gradient-based
  localization, in: Proceedings of the IEEE international conference on
  computer vision, 2017, pp. 618--626.

\bibitem{vaswani2017attention}
A.~Vaswani, N.~Shazeer, N.~Parmar, J.~Uszkoreit, L.~Jones, A.~N. Gomez,
  {\L}.~Kaiser, I.~Polosukhin, Attention is all you need, Advances in neural
  information processing systems 30.

\bibitem{ke2017lightgbm}
G.~Ke, Q.~Meng, T.~Finley, T.~Wang, W.~Chen, W.~Ma, Q.~Ye, T.-Y. Liu, Lightgbm:
  A highly efficient gradient boosting decision tree, in: Advances in neural
  information processing systems, 2017, pp. 3146--3154.

\bibitem{brier1950verification}
G.~W. Brier, Verification of forecasts expressed in terms of probability,
  Monthly weather review 78~(1) (1950) 1--3.

\bibitem{delong1988comparing}
E.~R. DeLong, D.~M. DeLong, D.~L. Clarke-Pearson, Comparing the areas under two
  or more correlated receiver operating characteristic curves: a nonparametric
  approach, Biometrics (1988) 837--845.

\bibitem{NIPS2017}
S.~M. Lundberg, S.-I. Lee, A unified approach to interpreting model
  predictions, in: Advances in Neural Information Processing Systems 30, 2017,
  pp. 4765--4774.

\bibitem{tiulpin2019multimodal}
A.~Tiulpin, S.~Klein, S.~Bierma-Zeinstra, J.~Thevenot, E.~Rahtu, J.~van Meurs,
  E.~H. Oei, S.~Saarakkala, Multimodal machine learning-based knee
  osteoarthritis progression prediction from plain radiographs and clinical
  data, arXiv preprint arXiv:1904.06236.

\bibitem{niu2021review}
Z.~Niu, G.~Zhong, H.~Yu, A review on the attention mechanism of deep learning,
  Neurocomputing 452 (2021) 48--62.

\end{thebibliography}

\newpage
% \processdelayedfloats
\clearpage
\makeatletter
% \efloat@restorefloats
\makeatother

\beginsupplement
\section*{Supplementary Material}
\subsection*{\textbf{Data Selection}}

\begin{figure}[!ht]
\centering
\includegraphics[width = 0.7\linewidth, ]{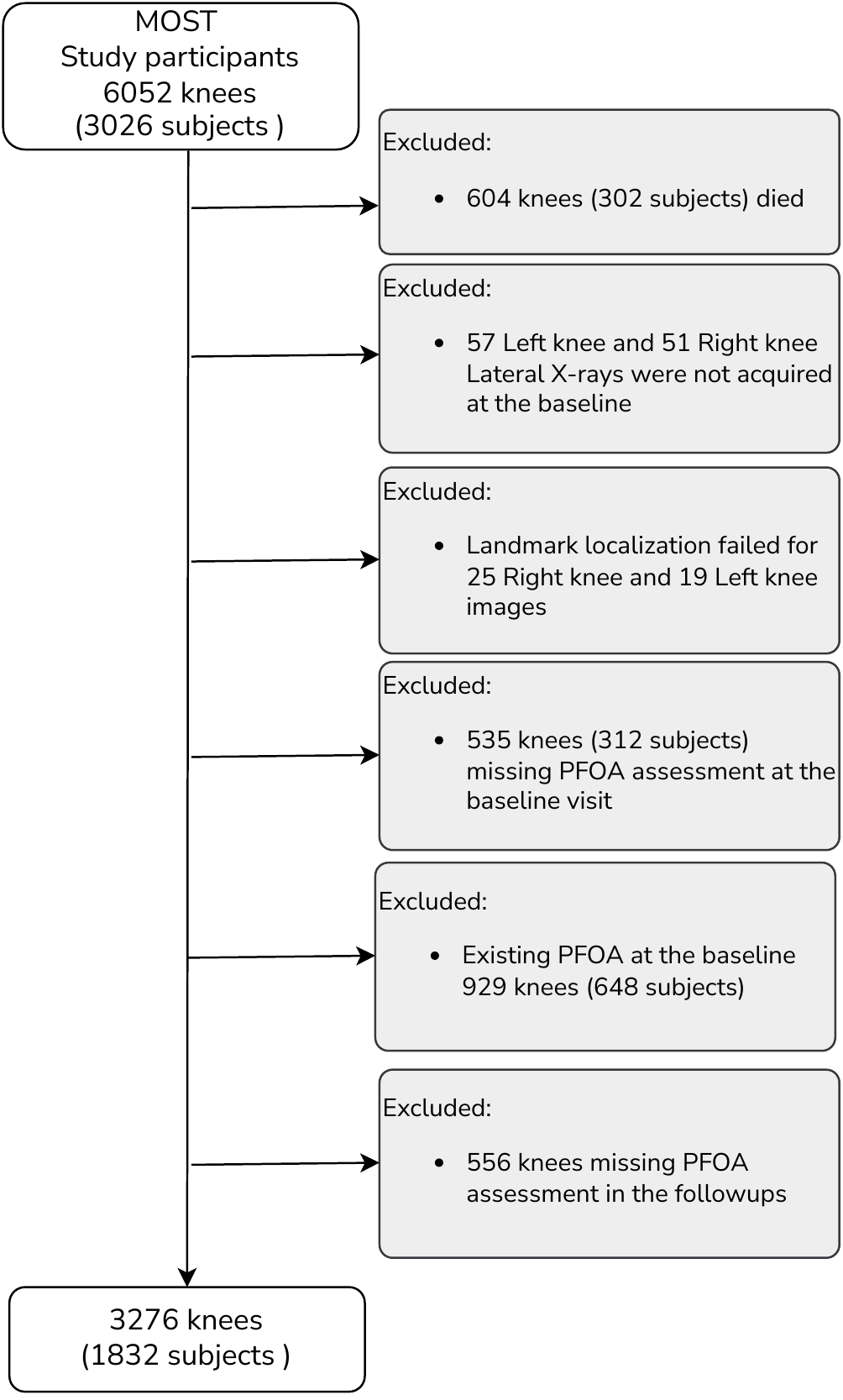}
\caption{Chart shows the selection of knees from the MOST study used in this work}
\label{fig:dataseletion}
\end{figure}

\clearpage
\subsection*{\textbf{Demographics}}
\begin{table}[hbt!]
\small
{\renewcommand{\arraystretch}{0.7}
    \caption{Demographics of the data used in this study (subset of MOST)}
    \label{tab:demographics}

\centerline{
\begin{tabular}{lrrrrrrrrrr}
\toprule
{} & {} & {} & \multicolumn{2}{l}{AGE} & \multicolumn{2}{l}{BMI} & \multicolumn{2}{l}{WOMAC} & \multicolumn{2}{l}{KL} \\
{PFOA} & \# of Females & \ \# of Males & mean & std & mean & std &  mean &  std & mean & std \\
% PFprog &      &     &      &     &       &      &      &     \\
\midrule
Non-Progressor &   1665 (58\%)  & 1208 (42\%) & 61.2 & 7.7 & 29.6 & 5.2 &  14.6 & 14.8 &  0.7 & 1.0 \\
Progressor    & 281 (70\%) & 122 (30\%) & 61.1 & 7.6 & 32.8 & 6.3 &  24.1 & 17.1 &  1.7 & 1.3 \\
\bottomrule
\end{tabular}
}
    }
\end{table}

\clearpage
\subsection*{\textbf{Feature Importance}}

\begin{figure}[hbt!]
\centering
\subfloat[Model1]{\includegraphics[width = 0.8\linewidth, ]{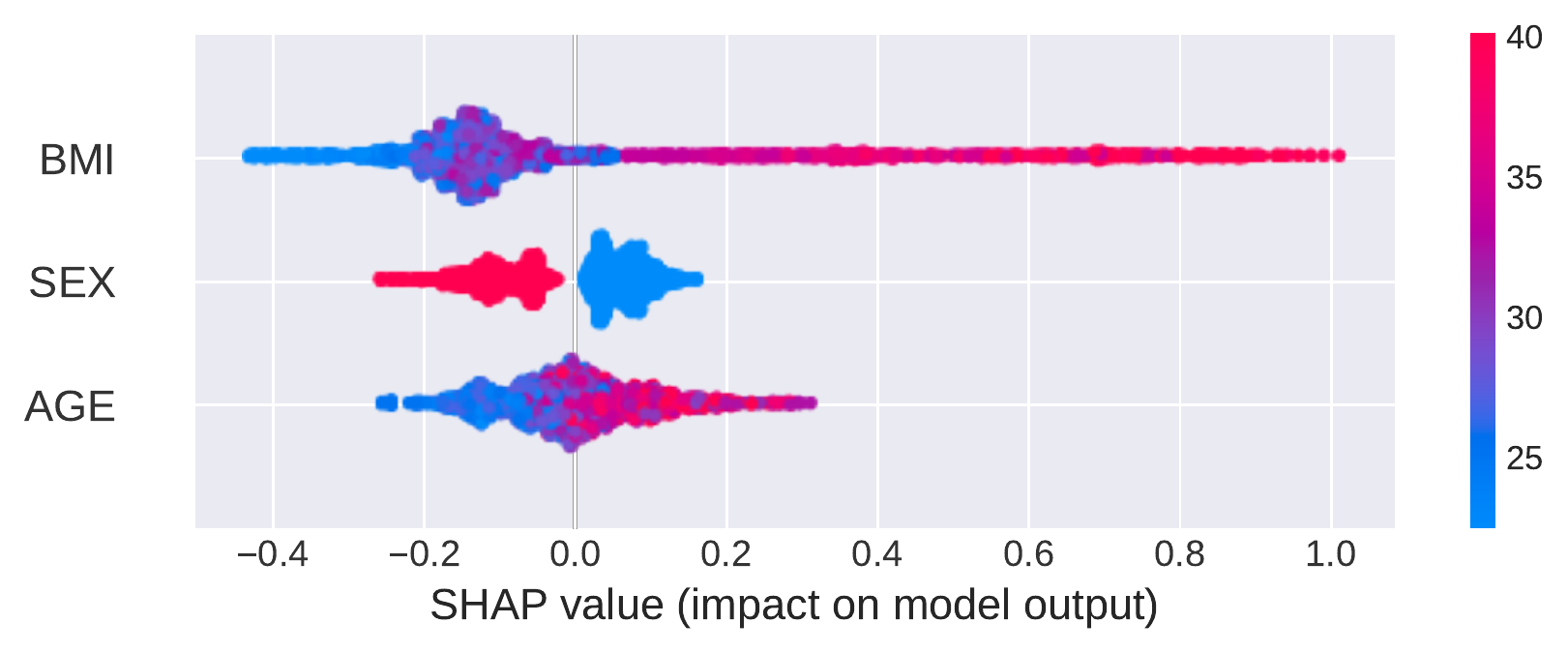}}\\
\subfloat[Model2]{\includegraphics[width = 0.8\linewidth, ]{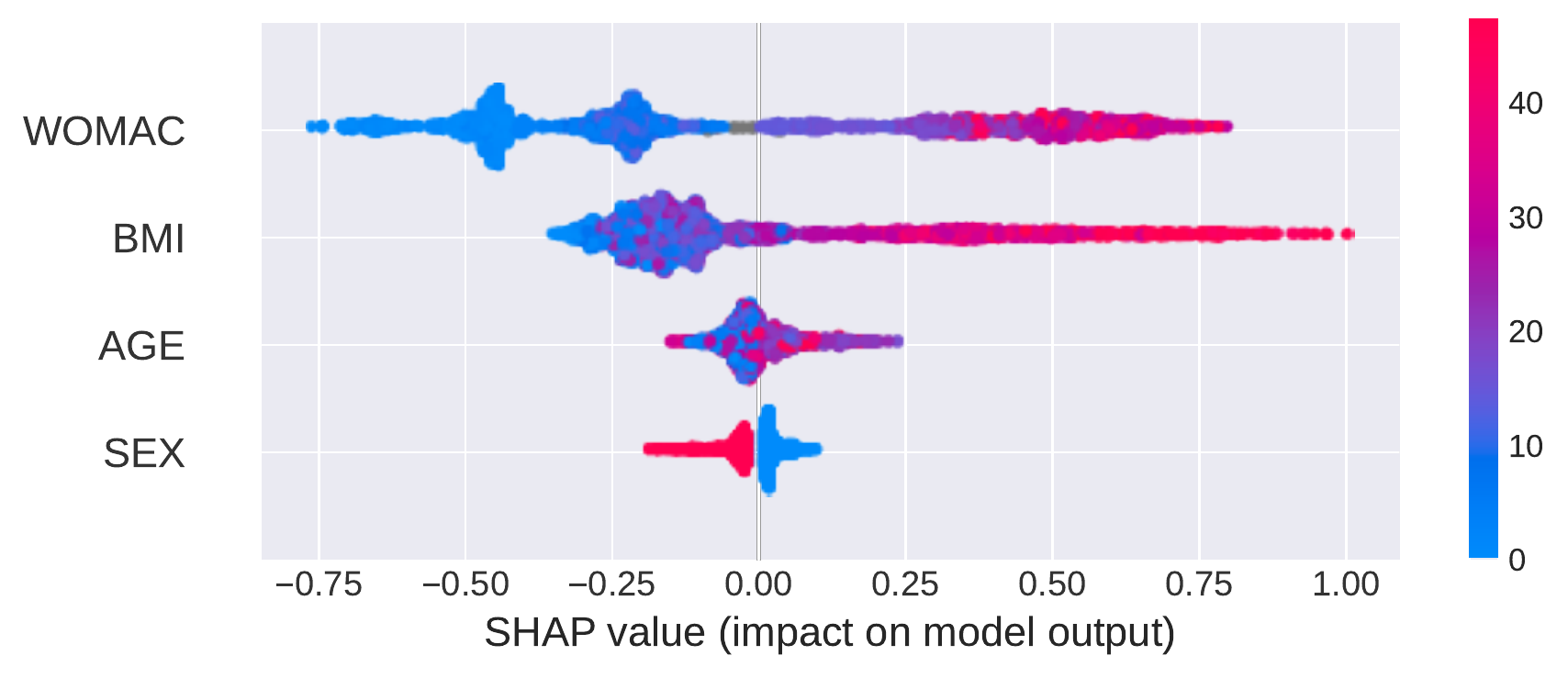}}\\
\subfloat[Model3]{\includegraphics[width = 0.8\linewidth, ]{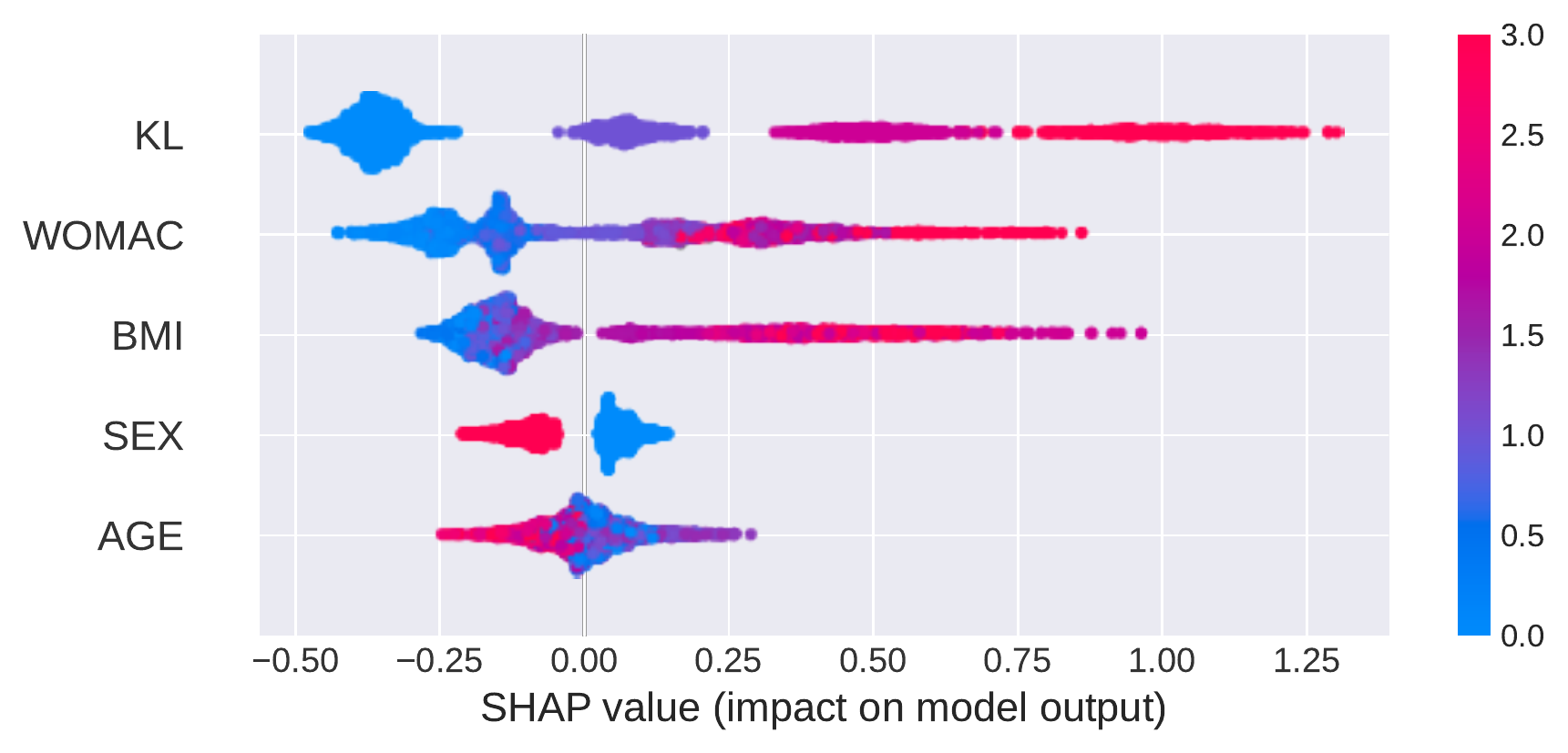}}\\
\caption{SHAP summary plot. High BMI, WOMAC and KL scores increase the predicted PFOA progression risk and low BMI, WOMAC, KL scores reduce the risk. }
\label{fig:sum}
\end{figure}

SHAP (SHapley Additive exPlanations) \cite{NIPS2017} is a method for explaining the output of any machine learning model.
It is a way to understand which features are driving the predictions of a model. It measures the impact of each feature on the model's output by looking at the contribution of that feature to the predicted outcome compared to what the predicted outcome would be if the feature was not included in the model. 
We also included summary where the plot shows the contribution of each feature to the model's output for each instance \ref{fig:sum}.
The SHAP feature importance values can be positive or negative, indicating whether the feature has a positive or negative impact on the prediction.
The instances are shown as dots along the x-axis, with jitter applied in the direction of y-axis for overlapping points to separate them visually. The dots are color-coded to indicate the value of the feature for that instance.

\clearpage
\subsection*{\textbf{Deep CNN Model and Attention Module}}

The VGG-16 model consists of 16 convolutional layers followed by three fully connected layers, and it achieved state-of-the-art performance on the ImageNet dataset at the time of its introduction.
Attention modules are a type of mechanism that can be added to neural networks to selectively emphasize or suppress certain parts of the input. These modules have been shown to be effective in improving the performance of deep learning models in various applications, including image classification, object detection, and machine translation.

\begin{figure}[hbt!]
\centering
\includegraphics[width = 1.1\linewidth, ]{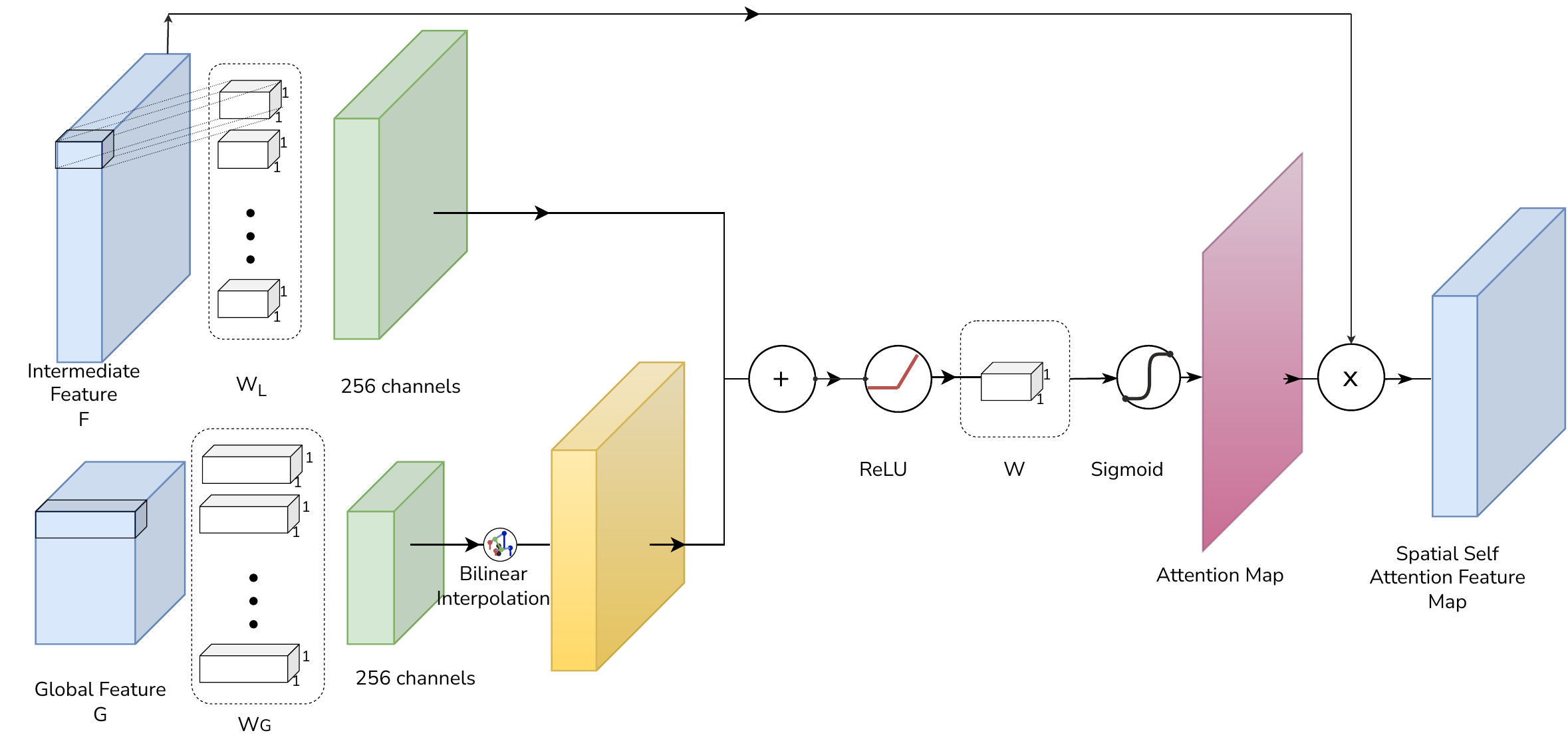}
\caption{Structure of the spatial attention module.}
\label{fig:attention}
\end{figure}

The structure of the attention module used in our study is shown in Figure \ref{fig:attention} \cite{yan2019melanoma}.
Spatial self attention map ($A$) is computed as follows. Let $F$ denotes an intermediate feature vector extracted at a given convolutional layer and $G$ denotes the global feature from the last convolutional layer. 

$A =    Sigmoid(  W \otimes ReLU ((W_L \otimes F) \oplus upsample(W_G\otimes G)) )$

The feature map with spatial attention ($\hat{F}$ ) is then computed by multiplying attention map ($A$) with the initial feature map ($F$).
If needed, global feature map is up-scaled using bi-linear interpolation to match the shape of the intermediate feature.
The final feature vector is then obtained by by concatenating the global average pooling of attention features and global feature $G$.

\clearpage
\subsection*{\textbf{Examples of Attention Maps}}

\begin{figure}[hbt!]
\centering
\includegraphics[width = 0.95\linewidth, ]{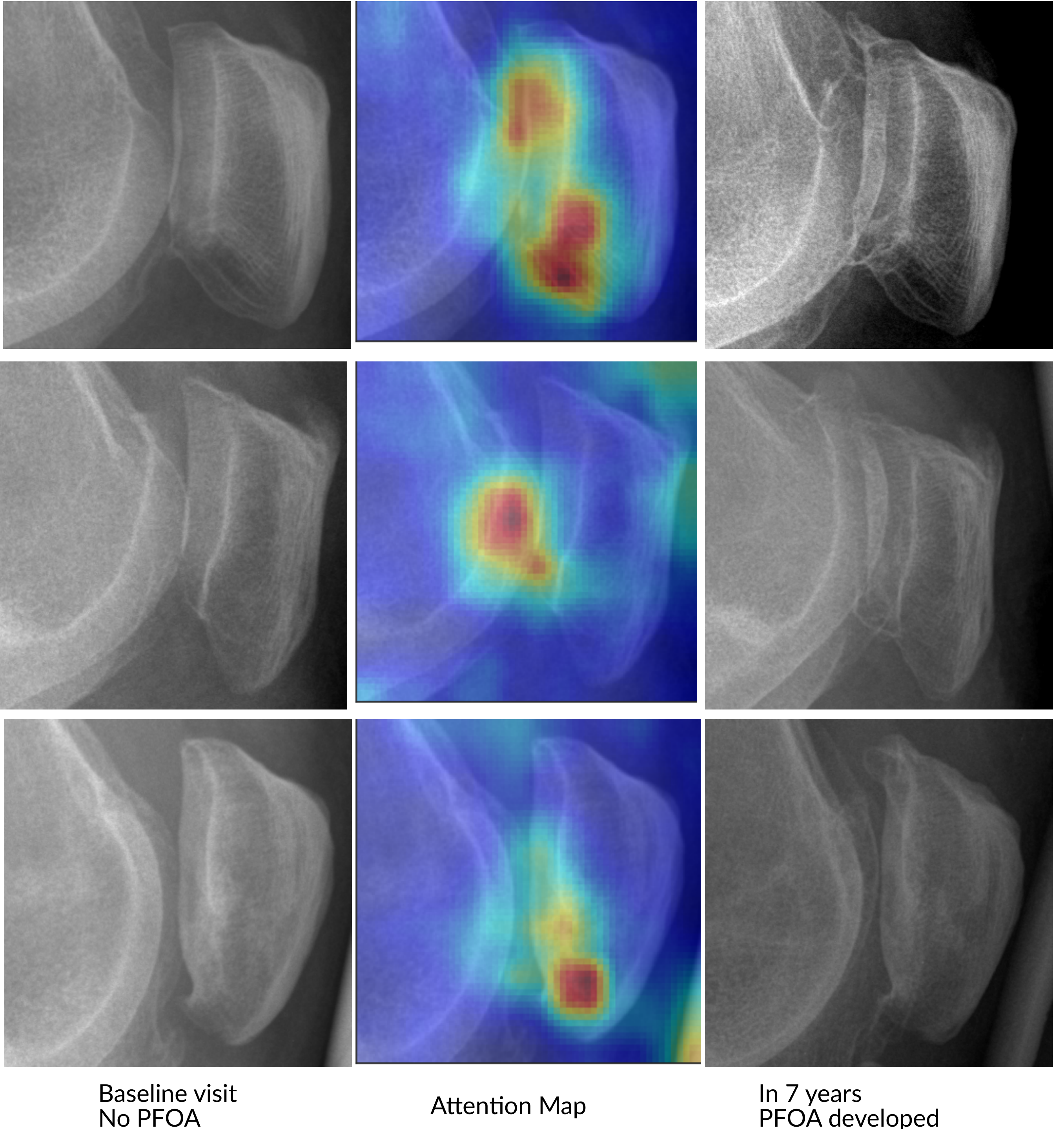}
\caption{Examples of trainable attention maps for the knees without baseline osteoarthritis and that progressed withing the next 7 years.}
\label{fig:maps1}
\end{figure}

\begin{figure}[hbt!]
\centering
\includegraphics[width = 0.95\linewidth, ]{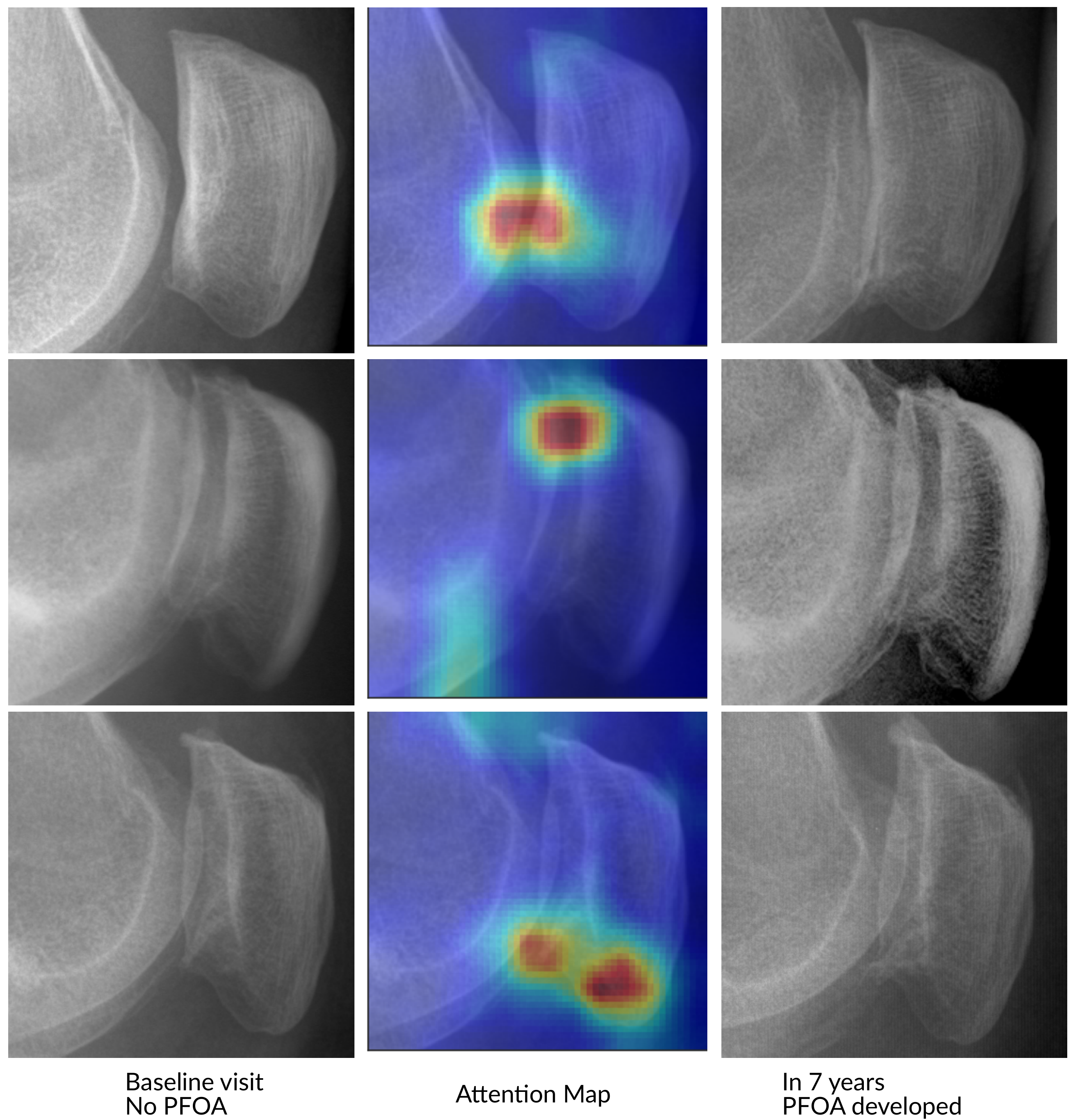}
\caption{Examples of trainable attention maps for the knees without baseline osteoarthritis and that progressed withing the next 7 years.}
\label{fig:maps2}
\end{figure}

\end{document}